\documentclass[12pt]{article}
\usepackage[totalheight = 23cm, totalwidth = 17cm]{geometry}

\usepackage{amsmath, amssymb, graphics, epsfig, graphicx}
\usepackage{dcolumn}
\usepackage{bm}
\usepackage{hyperref}
\usepackage{epstopdf}

\usepackage{color}
\usepackage[section]{placeins}

\newcommand{\nc}{\newcommand}
\nc{\ba}{\begin{eqnarray}}
\nc{\ea}{\end{eqnarray}}

\nc{\bfk}{{\bf k }}
\nc{\bfx}{{\bf x }}
  
\nc{\calP}{  { \cal P} }  
\nc{\calR}{  { \cal R} }  
\nc{\im}{ \mathrm{Im} }
  
\begin{document}

\begin{titlepage}


\begin{center}

\vskip 1.0cm

\large{\bf  Statistical Anisotropies in Gravitational Waves in Solid Inflation}

\vskip 1cm

\large{
Mohammad Akhshik$^a$,
\hspace{0.1cm}
Razieh Emami$^b$,
\hspace{0.1cm}
Hassan Firouzjahi$^{c}$, 
Yi Wang$^{d}$
}

\vskip 0.5cm

\small{\it 

$^{a}$ Department of Physics, Sharif University of Technology, Tehran, Iran  \\ 
$^{b}$ School of Physics, Institute for Research in Fundamental Sciences (IPM) \\
P.~O.~Box 19395-5531, Tehran, Iran\\
$^{c}$ School of Astronomy, Institute for Research in Fundamental Sciences (IPM) \\
P.~O.~Box 19395-5531, Tehran, Iran \\
$^d$Centre for Theoretical Cosmology, DAMTP, University of Cambridge, Cambridge CB3 0WA, UK
}

\vskip 1.2cm

\end{center}

\begin{abstract}

Solid inflation can support a long period of anisotropic inflation.
We calculate the statistical anisotropies in the scalar and tensor power spectra and their cross-correlation in anisotropic solid inflation. 
The tensor-scalar cross-correlation can either be positive or negative, which impacts the statistical anisotropies of the TT and TB spectra in  CMB map more significantly compared with the tensor self-correlation.
The tensor power spectrum contains potentially comparable contributions from quadrupole and octopole angular patterns, which is different from the power spectra of scalar, the cross-correlation or the scalar bispectrum, where the quadrupole type statistical anisotropy dominates over octopole.

\end{abstract}

\end{titlepage}

\setcounter{page}{0}
\newpage
\setcounter{page}{1}


\section{Introduction}
Cosmological observations from Cosmic Microwave Background (CMB) anisotropies strongly support inflation as the leading paradigm for the early universe 
\cite{Ade:2013lta, Ade:2013uln}. The recent detection of B-mode polarization in the CMB map 
 by the BICEP2 observation \cite{Ade:2014xna} has put inflation on an even firmer ground. This detection implies the existence  of primordial gravitational waves (GW), which is consistent with the simplest model of inflation. 
In its simplest realization, inflation is driven by a single scalar field slowly rolling over an approximately flat potential. The basic predictions of inflation in this simple realization is that the primordial perturbations on CMB are nearly scale-invariant, nearly adiabatic and nearly Gaussian.
In addition, depending on model parameters, namely if the inflaton field is super-Planckian,  GW with observable amplitude is generated. These generic predictions are very well consistent with recent observations \cite{Ade:2013lta, Ade:2013uln, Ade:2014xna}.

The detection of B-mode polarization with the implied amplitude of tensor-to-scalar ratio  $r= 0.2^{\, +0.07}_{\, -0.05}$ has triggered considerable interests in literature. One particular puzzle is that the 
implied amplitude of GW from BICEP2 observation is in some tension with the upper bound $r < 0.13$ from the Planck data. This tension may be alleviated after the Planck observation releases its polarization data. 
In addition, the forthcoming observations such as SPTPol \cite{Austermann12}, ACTPol \cite{Niemack10}, POLARBEAR \cite{PolarBear} and CLASS \cite{Eimer12} can not only confirm the BICEP2 detection but also search for more refined features of the primordial B-mode polarization and GW. Therefore, models of inflation with non-trivial features  can be  directly compared with the polarization data.  Particularly interestingly, there are models which predict statistical anisotropies in scalar and GW power spectra.  Recently primordial statistical anisotropies in GW in models of anisotropic inflation have been studied in \cite{Chen:2014eua}. It is argued that the non-trivial anisotropic features in GW  power spectrum may be behind various low-$\ell$ anomalies in CMB map. This is particularly interesting, since the contribution of GW in temperature power spectrum dies off rapidly for
high $\ell$ so GW with non-trivial features may be behind the various  anomalies 
which have been observed in the low-$\ell$ CMB map only.

With the above motivation in mind, in this paper we study scalar and tensor perturbations in the model of anisotropic solid inflation.  Solid inflation \cite{Endlich:2012pz} is a novel model of inflation which has a number of interesting properties both at the level of background and at the level of perturbations. Here inflation is driven by 
a configuration which resembles a solid. In this picture the space may be fragmented into small cells in which the location of each cell is collectively defined by the scalar fields $\phi^I$ for $I=1, 2$ and $3$. At the background level, the position of each cell is given by
\ba
\label{back-phi}
\langle \phi^I \rangle = x^I   \quad , \quad I=1, 2, 3.
\ea
This is a peculiar property of solid inflation in which the scalar fields $\phi^I$ are time-independent at the background level. Having this said, the ansatz (\ref{back-phi}) naively seems to violate the isotropy and the homogeneity of the cosmological background. In order to keep the background isotropic and homogeneous, the following internal symmetries on the matter fields Lagrangian are imposed
\ba
\label{cond1}
\phi^{I} \rightarrow \phi^I + C^I
\ea
and
\ba
\label{cond2}
\phi^I \rightarrow O^I_J \phi^J \quad , \quad O^I_J \in SO(3)
\ea
in which $C^I$ are constants and $O^I_J$ belong to $SO(3)$ rotation group. 
The translation invariance in field space, Eq. (\ref{cond1}),  implies that the dynamical quantities in the Lagrangian are the derivatives of the scalar fields $\partial \phi^I$. As a result the background ansatz, Eq. (\ref{cond1}), is consistent with the translational invariance of the cosmological background. In addition, the internal $SO(3)$ rotation invariance ensures that the background is also isotropic. Therefore, equipped with the internal symmetries (\ref{cond1})
and (\ref{cond2}), the background expansion is consistent with the cosmological principles. 
Further insights on the properties of solid inflation can be found at \cite{Endlich:2012pz}.

Perturbations in solid inflation show interesting features. First, although the model 
looks like a three-field inflationary models at the background level, but as far as the scalar 
perturbations are concerned it is effectively a single field model. The scalar perturbations are described by a single field $\pi$, dubbed as ``phonons" in \cite{Endlich:2012pz}, corresponding to the perturbations of the overall  volume of the solid. Second, large non-Gaussianities are generated, with the shape similar to the local shape. This is in contrast to the celebrated Maldacena's consistency condition for single field models of inflation in which the amplitude of local-like non-Gaussianity $f_{NL}$ is  at the order of slow-roll parameters \cite{Maldacena:2002vr}. 

Due to the solid nature of the Lagrangian, the model is very inefficient in erasing classical anisotropies. As a result, as noticed in \cite{Bartolo:2013msa}, solid inflation naturally sustains a long period of anisotropic inflation in which the background is in the form of Bianchi I universe. This model highly
resembles the models of anisotropic inflation with a background gauge field. As studied in 
\cite{Chen:2014eua, Watanabe:2010bu} scalar and tensor perturbations in models of anisotropic inflation with the background gauge fields have statistical anisotropies which may be imprinted on the CMB polarization maps. In particular, it is emphasized in \cite{Chen:2014eua} that the GW may be a more sensitive probe of statistical anisotropies. With this motivation, in this paper we study the imprints of statistical anisotropies in GW in the model of anisotropic solid  inflation. Note that the statistical anisotropies in curvature perturbation power spectrum in the model of anisotropic solid inflation  were studied in  \cite{Bartolo:2013msa, Endlich:2013jia}. Here we go further and look at statistical anisotropies in GW and the scalar-tensor cross-correlation.

 
\section{Solid Inflation}
\label{review}

In this Section we briefly review solid inflation, further details can be found at the original paper 
\cite{Endlich:2012pz}.  In order to get insights about the model here we first 
present the isotropic solid inflation model. Then we move on to the anisotropic solid inflation setup
which is our model of interest in this work.

\subsection{The Background}
\label{back-sec}

As described in the Introduction Section, the model consists of three scalar field $\phi^I$ with the internal symmetries  (\ref{cond1}) and (\ref{cond2}) in order to obtain an isotropic and homogeneous cosmological background. The most general action consistent with these symmetries coupled minimally to gravity is given by 
\ba
\label{solid-action}
S = \int d^4 x \sqrt{-g} \left\{ \frac{M_P^2}{2} R + F[X, Y, Z] 
\right\} \, ,
\ea
 in which $M_P =1/8 \pi G$ is the reduced Planck mass with $G$ being the Newton constant 
 and $F$ is a function encoding the properties of the solid. The variables $X, Y$ and $Z$ are functions of the derivatives of $\phi^I$. Noting that the action is invariant under the internal symmetries  (\ref{cond1}) and (\ref{cond2}), then
 $F$ is a function of the $SO(3)$ invariant matrix $B^{IJ}$ in which
 \ba
 \label{B-def}
 B^{IJ} \equiv g^{\mu \nu} \partial_\mu \phi^I \partial_\nu \phi^J \, .
 \ea
Note that we choose the convention in which the Greek indices $\mu, \nu, ...$ represents the four-dimensional spatial coordinates while the capital Latin indices $I, J, ...$ stand for the three-dimensional 
internal matter field space. 
To determine the functional form of $F$, one has to construct scalars from the matrix $B^{IJ}$. For a $3\times 3$ matrix, the three independent options are 
\ba
\label{XYZ-def}
X \equiv  [B]  \quad , \quad
Y \equiv \frac{[B^2]}{ [B]^2} \quad , \quad
Z\equiv \frac{ [B^3]}{  [B]^3}  \, ,
\ea
in which $[ B]  \equiv  \mathrm{Tr}B = B^{I I}$ where we have used the convention that the doubly repeated 
indices are summed over. Also note that the internal indices $I, J$ are raised and lowered by
the Euclidean metric $\delta_{IJ} $.  Note that the variables $Y$ and $Z$ are defined such that 
they are insensitive to the overall scaling of the volume so only the variable $X$ controls the overall volume of the space. 

The background space-time metric is given by the usual flat FRW metric 
\ba
ds^2  = -dt^2 + a(t)^2 d \bfx^2
\ea
in which $a(t) $ is the background scale factor.  At the background level, one can show that
\ba
X= \frac{3}{a(t)^2} \quad , \quad Y= \frac{1}{3}  \quad , \quad Z=\frac{1}{9} \, .
\ea
As explained before, $Y$ and $Z$ are defined such that they are insensitive to the volume of space while the information about the background volume is entirely encoded in $X$. Also note that  $a(t)$ is not physical by itself because one can re-scale it by a constant factor and absorb it into the 
comoving coordinate $x^i$. As a result, $X=3/a^2$   is not physical by itself either. This is understood from  the background ansatz $\langle \phi^I \rangle = x^I$ which implies that 
$\phi^I$ and $B^{IJ}$, from which $X$ is made of,  are not physical. We will discuss later on as how one can define physical observable and the physical clock.

The energy momentum-tensor for the Einstein equations are
\ba
\label{T-munu}
T_{\mu \nu} &=& g_{\mu \nu} F - 2 \partial_\mu^I \partial_\nu \phi^J \frac{\partial F}{\partial B^{IJ}} \nonumber\\
 &=&   g_{\mu \nu} F  - 2 \partial_\mu^I \partial_\nu \phi^J\left[  \left( F_X - \frac{2 Y F_Y}{X} - \frac{3 Z F_Z}{X}  \right) \delta^{I J} + \frac{2 F_Y B^{IJ}}{X^2} + 
 \frac{3 F_Z B^{IK} B^{KJ}}{X^3}     \right]
\ea
in which $F_X \equiv \partial F/\partial X$ and so on. 

The background cosmological equations are
\ba
\label{back-eq}
3 M_P^2 H^2 = \rho \quad , \quad  \dot H = -\frac{1}{2 M_P^2} (\rho + p) \, .
\ea
in which $H= \dot a(t)/a(t)$ is the Hubble expansion rate. 
For the $T_{\mu \nu}$ given in Eq. (\ref{T-munu}), the energy density $\rho$ and the pressure $p$ are given by
\ba
\label{rho-p}
\rho = -F \quad , \quad p= F- \frac{2}{a^2} F_X \, .
\ea
Finally,  varying the action with respect to $\phi^I$ yields the scalar field equations
\ba
\label{KG-eq}
\partial_\mu \left( \sqrt{-g} \frac{\partial F}{\partial B^{ab}} \frac{\partial B^{ab}}{\partial \partial_\mu \phi^I} \right) =0  \, .
\ea
The interesting observation is that at the background level $\phi^I$ are independent of $t$
and the scalar field equations are satisfied automatically. As a result, we do not get any information from Eq. (\ref{KG-eq}) at the background level.  

To obtain  a long enough period of inflation, we need the slow-roll conditions to be satisfied. Defining the slow-roll parameters via
\ba
\label{epsilon-eta}
\epsilon \equiv - \frac{\dot H}{H^2} \quad , \quad \eta = \frac{\dot \epsilon_H }{H\epsilon_H } \, ,
\ea
we need $\epsilon \ll 1$ and $\eta \ll 1$ during inflation.  For the solid model, the slow-roll parameters are calculated to be
\ba
\epsilon = \frac{X F_X}{F} \quad , \quad 
\eta = 2\left( \epsilon -1 - \frac{X^2 F_{XX}}{X F_X} \right) \, .
\ea
To satisfy the  condition $\epsilon \ll 1$ we require $F$ to have a very weak dependence in X.  Physically, this means that the dominant source of energy to drive inflation comes from a cosmological constant term. Note that although we follow the usual convention in calling $\epsilon $ and $\eta $ as the slow-roll parameters, however in solid inflation nothing is rolling. Indeed, at the background level $\phi^I$ are exactly time-independent so they do not roll towards a minimum as usually assumed in conventional models of inflation. In a sense, the slow-roll assumption here means that the physical parameters such as $H$ evolves slowly and have a very weak time-dependence during inflation.

Finally, one may wonder how inflation ends in this setup  and how the universe reheats  after inflation.  These are somewhat open  questions in solid inflation. To answer these questions qualitatively let us look at the question what is the physical clock of the system? 
As discussed before, neither $a(t)$ nor $X$ are physically observable. However, $F$ which has the dimension of $M^4$ is physical since it determines $H$ via $3 M_P^2 H^2 = -F$. Therefore, the value of  $F$ is a good candidate for the physical clock of the system, i.e. $F^{-1/4}$ is a measure of time. This is equivalent to the assumption that  $\rho$ or $p$ are
good candidates for  the physical clock as envisaged in \cite{Endlich:2012pz}. As a suggestion, one can imagine that in this setup inflation ends when the value of $F$ reaches a  critical value, $F_e$, in which a rapid phase transition occurs terminating inflation abruptly. 
This is somewhat similar to models of hybrid inflation in which inflation is terminated by a sharp waterfall phase transition once the inflaton field reaches a critical value. For this picture to work one needs to couple the solid fields to the additional dynamical fields to trigger the rapid phase transition. Furthermore, in order not
to affect the super-horizon curvature perturbations we assume that the phase transition from a solid to a radiation-dominated era happens quickly. Finally, to reheat the universe, we assume that the solid model is coupled to additional light scalar and gauge fields so the energy in solid is dumped into Standard Model degrees of freedom via the phase transition. Having all these said, whether or not inflation can be terminated and the universe reheats successfully afterwards have to be studied in details which are beyond the scope of this work.

\subsection{Perturbations in Solid Inflation}
\label{power-sub}

In this subsection we briefly review perturbation analysis in solid inflation, for further details see   \cite{Endlich:2012pz} and \cite{Bartolo:2013msa}. 

The scalar fields perturbations are given by
\ba
\phi^I = x^I + \pi^I(t, \bfx) \,.
\ea
In addition, we can decompose the perturbations $ \pi^I(t, \bfx)$ into the transverse part $\pi_T$ and  the longitudinal part $\pi_L$ via \footnote{Note that our convention in 
Eq. (\ref{pi-decompose}) is different than the convention used in \cite{Endlich:2012pz} 
in which it is assumed $\pi^i(t, \bfx) = \frac{\partial_i}{\sqrt{-\nabla^2} } \pi_L(t, \bfx) + \pi^i_T(t, \bfx)$. When going to the Fourier space this brings an additional factor $k$ compared to \cite{Endlich:2012pz} .}
\ba
\label{pi-decompose}
\pi^i(t, \bfx) = {\partial_i} \pi_L(t, \bfx) + \pi^i_T(t, \bfx)
\ea 
subject to the condition $\partial_i \pi^i_T =0$.  Note that from now on we do not distinguish between the spatial indices $i, j, ...$ and the internal indices $I, J, ...$.

As argued in  \cite{Endlich:2012pz} $\pi_L(t, \bfx)$ plays the role of ``phonons'' for the longitudinal 
fluctuations of the solid.  The sound speed of the longitudinal excitations or phonons is given via
\ba
c_L^2  \equiv  1+ \frac{2F_{XX} X^2}{3F_X X} + \frac{8(F_Y + F_Z)}{9 F_X X} \, .
\ea
In addition,  the sound speed associated with the transverse excitations $c_T$ is given by 
\ba
c_T^2 = 1 + \frac{2(F_Y + F_Z)}{3 X F_X} =  
\frac{3}{4}(1+ c_L^2 - \frac{2\epsilon}{3} + \frac{\eta}{3}) \, .
\ea
As is clear from the above formulas, the combination $F_Y + F_Z$ plays important roles in perturbation theory. Requiring that $c_T$ to be sub-luminal we obtain $(F_Y + F_Z)/\epsilon F  \leq 0$. In addition, requiring both transverse and longitudinal modes to be stable  (i. e. non-tachyonic with $c_T^2, c_L^2 >0$) we need 
$(F_Y + F_Z)/\epsilon F  \ge -1/72 + \eta/72 -\epsilon/36$. As a result, discarding the small slow-roll corrections, we are left with the small window 
\ba
\label{FYZ-cond}
-\frac{3}{8} \leq \frac{F_Y + F_Z}{\epsilon F}  \leq  0  \, .
\ea

One peculiar property of solid inflation model is that  $\calR$ or $\zeta$, corresponding respectively to curvature perturbations on comoving slices and uniform energy slices, are time-dependent on super-horizon scales. As a result, in general $\zeta \neq -\calR$ on super-horizon scales. 
However, this time-dependent corrections are at the order of $\epsilon$ which be discarded at leading order when we consider small anisotropy limit in our analysis in next Section. In this limit we take $\calR =-\zeta$ as in simple models of inflation. 

In flat gauge $\zeta$ is given by $
\zeta = - H \frac{\delta \rho}{\dot \rho}= \frac{\delta \rho}{3 (\rho+p)} $. 
On the other hand, in solid inflation $\rho+ p =2 F_X/a^2$ and $\delta \rho = -F_X \delta X$.
Using $\delta X = 2 \partial_i \delta \phi^i/a^2 = -(2 k^2/a^2) \pi_L$, in Fourier space we obtain 
\ba
\label{zeta-pi} \zeta = -\frac{k^2}{3} \pi_L \, .
\ea
This indicates that at the linear order one can work with either $\pi_L$ or $\zeta$. 

The wave function of $\zeta$ to leading order in slow-roll parameters is 
\ba
\label{zeta}
\zeta = -\frac{C}{3 c_L^2}  \left(- k c_L \eta \right)^{\frac{5}{2}  } H_{ \frac{5}{2} }^{(1)}( - k c_L \eta  )
\ea
in which $\eta$ is the conformal time related to the cosmic time via $d\eta = d t/a(t)$ and 
the normalization constant $C$ is given by 
\ba
C = \frac{-i \sqrt{\pi} H}{2 M_P \sqrt{2 \epsilon k^3 c_L}} \, .
\ea
The curvature perturbations power spectrum at the end of inflation $\eta_e$ is
\ba
\langle \zeta_{\bfk_1} \zeta_{\bfk_2}^* \rangle  =  (2\pi)^3 \delta ^3(\mathbf{k_1}+\mathbf{k_2})
P_\zeta(k_1) 
\ea
with
\ba
\label{power}
P_\zeta(k) = | \zeta_k(\eta_e)|^2 \simeq 
\frac{H^2 }{4 \epsilon c_L^5 M_P^2 k^3} \quad , \quad 
\calP(k) \equiv \frac{k^3}{2 \pi^2} P_\zeta(k)
\ea

Now considering the tensor perturbations, the wave function of the two polarizations $s=+, \times$ of
the tensor perturbations is given by (we will present more details of the convention for the polarization tensor in next Section when studying anisotropic background)  
\ba
\label{hs-wave}
{h}_{s}(k,\eta) = \frac{2 i H\eta }{M_{P}\sqrt{2 k}}\left(1-\frac{i}{ k\eta} \right)e^{-ik\eta} ~~~,~~~ (s= +, \times) \, .
\ea
The power spectrum of the tensor perturbations $\calP_h(k) =\frac{ k^3  }{ 2 \pi^2} |h_s(k)|^2 $
is given by
\ba
\label{calPh0}
\calP_h^{(0)} = \frac{2 H^2}{\pi^2 M_P^2}  = 16 \epsilon  c_L^5 \, \calP_\zeta^{(0)} \, .
\ea
Therefore, defining the tensor-to scalar ratio $r \equiv \calP_h/\calP_\zeta  $ we have $r = 16 \epsilon c_L^5$ for the isotropic theory. Note the additional factor $c_L^5$ which appears in the parameter $r$ compared to conventional models of inflation. 

Before closing this subsection on solid inflation, there are few important remarks in order. First, one can show that at the linear level of perturbations $\pi_T$ couples to the vector parts of the metric perturbations. In addition, the wave function of 
$\pi_T$ is suppressed compared to that of $\pi_L$ by a factor $\epsilon$  \cite{Endlich:2012pz}. Therefore, the contributions 
of $\pi_T$ in anisotropies are suppressed compared to the contributions from $\zeta$ and $h_s$.
In addition, the vector perturbations are not supported in the isotropic background after inflation.
Therefore,  the contributions of $\pi_T$ in CMB anisotropies are sub-leading and 
we do not study $\pi_T$ in our analysis below. The second comment is that it is well-known that solid predicts a blue spectral tilt for tensor perturbations $n_T -1 \simeq 2 \epsilon c_L^2$.
This may have important implications in resolving the apparent tension between the Planck and the BICEP2 observations \cite{Wang:2014kqa, Brandenberger:2014faa}. Technically, those effects originates from the evolution of $h_{ij}$ on super-horizon scales (similar to evolutions of $\calR$ and $\zeta$ on super-horizon scales) which comes from the slow-roll corrections in the wave functions. However, in our analysis of anisotropies in next Section, we consider the wave function with the simple forms given in Eqs. (\ref{zeta}) and (\ref{hs-wave}) with no slow-roll corrections. The inclusion of the slow-roll corrections in wave functions bring the sub-leading corrections in our anisotropy analysis which can be discarded. Finally, as we mentioned, the  tensor-to scalar ratio
is given by $r = 16 \epsilon c_L^5$. For $\epsilon, \eta $ at the order few percents one usually 
gets $c_L^2 \sim 1/3$ and $r$ is very small. However, one can look for the parameter space of solid model in which $\epsilon, \eta$ and $s \equiv \dot c_L/c_L$ are not very small, perhaps at the 
order 5 to 10 percents. In this limit one can increase $c_L$ above $1/\sqrt3$ while both $c_L$ and $c_T$ being still sub-luminal. As a result (as we verified numerically) with some tunings on slow-roll parameters  there are corners of parameter space in solid which can lead to large enough
value of $r$, say $r=0.1$, which can be consistent with both Planck and BICEP2 data.

\subsection{Anisotropic Solid Inflation}

After reviewing the background and perturbations in solid inflation, now we look at anisotropic solid inflation. As studied in \cite{Bartolo:2013msa} solid is nearly insensitive to the spatial expansion so it is not efficient in erasing anisotropic deformation of the background geometry. As a result, a long period of anisotropic inflation is achieved. This should be compared with the other known mechanism of obtaining anisotropic inflation employing the $U(1)$ gauge field with the action $f(\phi)^2 F_{\mu \nu} F^{\mu \nu}$ in which $f(\phi)$ is appropriately chosen to break the conformal invariance \cite{Watanabe:2009ct, Emami:2010rm}. 

We consider the Bianchi I background with the metric 
\ba
\label{metric-0}
ds^2 =-dt^2 + a(t)^2 d x^2 + b(t)^2 \big( dy^2 + d z^2 \big)
\ea
with the identifications
\ba
a(t) \equiv e^{\alpha - 2 \sigma} \quad , \quad
b(t) \equiv e^{\alpha + \sigma} \, .
\ea
In this convention, $e^{\alpha}$ measures the average expansion while $\sigma$ is a measure
of anisotropies. In order to be consistent with the cosmological observation the background has to be
nearly isotropic so $\sigma \ll 1$. For example, one may assume $\sigma \sim \epsilon$.
Also note that we have assumed the residual symmetry in $y-z$ plane. 
In principle one can consider the background with no such residual symmetry. Finally, note that 
with the normalization used to set $\phi^I =x^I$ at the background level, there is no  freedom left to absorb $\sigma$ by a rescaling of $x, y$ and $z$ coordinates. Therefore, $\sigma$ is physical. This is in contrast to models of anisotropic inflation from the gauge field \cite{Watanabe:2009ct, Emami:2010rm} in which only $\dot \sigma$ is physical. 

The dynamics of the background anisotropic inflation was studied in \cite{Bartolo:2013msa}. The Background Einstein equations are
\ba
\dot \alpha^2 -\dot \sigma^2  &=& - \frac{F}{3 M_P^2} \, , \\
\ddot \alpha + 3 \dot \sigma^2  &=&  \frac{e^{4 \sigma} + 2 e^{-2 \sigma}}{3 M_P^2} e^{-2\alpha} F_X \, , \\
\ddot \sigma + 3 \dot \sigma \dot \alpha  &=& \frac{2 ( e^{4 \sigma} -  e^{-2 \sigma} )}{3 M_P^2}
e^{-2\alpha} F_X  - \frac{4 e^{6\sigma} ( 1-e^{6 \sigma}  ) F_Y}{(2+ e^{6 \sigma})^3 M_P^2}
- \frac{6 e^{6\sigma} (1-  e^{12 \sigma})  F_Z }{( 2+ e^{6 \sigma}  )^4 M_P^2 } \, .
\ea
In the small anisotropy limit $\sigma \ll 1$ the last equation above reduces to
\ba
\ddot \sigma + 3 H \dot \sigma + 4 \epsilon H^2 c_T^2 \sigma \simeq 0
\ea
in which $H \equiv \dot \alpha$ is the average Hubble expansion rate and $c_T$ is the speed of sound propagation for the transverse mode. For nearly constant values of $\epsilon $ and $c_T$ the above equation can be solved yielding \cite{Bartolo:2013msa}
\ba
\label{sigma-t}
\sigma(t) \simeq \sigma_1 e^{-\int dt\, \left[ (3- (2+c_L^2) \epsilon ) \right] H }
+ \sigma_2  e^{-\int dt\, \frac{4}{3} c_T^2 \epsilon H  }
\ea
in which $\sigma_1$ and $\sigma_2$ are two constants. The first solution above represents the 
fast decaying solution as in conventional models of inflation. 
The second solution represents the slow-decaying solution which we are looking for and is unique to the solid model. As argued before, this originated from the fact that the solid is not efficient 
in erasing the anisotropic deformation of the background imposed from the initial conditions. However, if inflation lasts long enough the anisotropy decays and one reaches the 
isotropic FRW solution. In other words, the FRW universe is the attractor solution of the solid background. Neglecting the running of $\epsilon $ and $c_T$, from Eq. (\ref{sigma-t}) we conclude that  the FRW attractor solution is reached if inflation lasts longer than $1/\sqrt{\epsilon} \, c_T$.


\section{Anisotropic Gravitational Waves}
\label{GW-anisotropy}

Having studied solid inflation and its anisotropic extension in the previous Section, now we study statistical anisotropies in scalar and tensor power spectra and their cross-correlation induced from the 
background anisotropy. 

To calculate the anisotropic power spectra we employ the perturbative method of in-in formalism. In this picture the free theory is given by
the isotropic solid inflation with non-interacting scalar and tensor perturbations respectively given by 
Eqs. (\ref{zeta}) and  (\ref{hs-wave}). Then we treat the change in Hamiltonian from anisotropy as the 
interaction Hamiltonian. This way, we can calculate the induced anisotropies in curvature power spectrum and the GW power spectrum to all orders in powers of $\sigma$ perturbatively. In addition, we get non-zero cross-correlation $\langle \zeta h_s \rangle$ induced from anisotropies.

To calculate the full interaction Hamiltonian   we have to perform the full metric perturbations including the scalar, vector and tensor perturbations 
along with the transverse and longitudinal perturbations of the matter sector $\pi_L$ and $\pi_T$.
In general this is a very complicated task. However, things become considerably simplified if we 
employ the experience with the similar situations in anisotropic inflation. It is shown in models 
of anisotropic inflation with gauge fields that the dominant contribution in interaction Hamiltonian come from the  matter sector while the contributions from the gravitational sectors are slow-roll suppressed \cite{Emami:2013bk, Bartolo:2012sd}. In this case, to calculate the leading order anisotropies, one can neglect the perturbations induced from the gravitational sector and only concentrate on perturbations induced from the matter sector Lagrangians. In particular, one does not need to consider the complicated process of eliminating the non-dynamical metric perturbations $\delta g_{0 \mu}$. Here we present our analysis of anisotropies considering only the perturbations originating from the matter sector. However, we have checked the perturbations from the whole gravitational and metric sectors including all perturbations in metric and $\delta \phi^I$. We have checked that indeed the leading source of anisotropies is generated from the matter sector.

With these discussions in mind, and considering the flat gauge in which $\zeta = -k^2 \pi_L/3$, we consider the metric perturbations  as follows 
\ba
\label{metric-pert}
ds^2 = - dt^2 + a_i(t) a_j(t) \bigg( \delta_{ij} + h_{ij} \bigg) dx^i dx^j
 \ea
in which $h_{ij}$ represents the tensor perturbations subject to the transverse and the traceless 
conditions  $h_{ii} = h_{ij,j} =0$. In this notation $a_i(t)$ represents either $a_x(t) =a(t) = e^{\alpha-2 \sigma}$  or $a_y(t) = a_z(t) = b(t)= e^{\alpha+ \sigma}$ as given in Eq. (\ref{metric-0}). 

Now we present our decomposition of the tensor perturbations  $h_{ij}$ into $h_\times$ and $h_+$ polarizations following the method of  \cite{Ohashi:2013qba} and \cite{Chen:2014eua}. 
Decomposing $h_{ij}$ into $e_{ij}^{(s)}(\bfk)$ in Fourier space and imposing  the traceless  and transverse conditions we get 
\ba
e_{i i}^{(s)}(\bfk) = 0 \quad , \quad  k_j e_{i j}^{(s)}(\bfk) = 0  \, ,
\ea
in which $s= \times, +$ represents the two polarization modes of the tensor perturbations. Our normalization is such that
\ba
e^{(s)}_{ij}(\mathbf{k}) e^{*(s')}_{ij}(\mathbf{k}) = \delta_{ss'} \, ,
\ea
where $*$ stands for  the complex-conjugation. In addition we also have $e^{(s)}_{ij}(\mathbf{k}) = e^{*(s)}_{ij}(\mathbf{-k})$.

The  quantum operators associated with $h_{ij}$ is represented by 
$\widehat{h}_{ij}(\mathbf{k},\eta) $ which  in terms  of the annihilation and creation operators 
are given by 
\ba
\label{tensor}
\widehat{h}_{ij}(\mathbf{k},\eta) = \sum _{s=+,\times}  \widehat{h}_{s}(\mathbf{k},\eta)
e_{ij}^{(s)}(\bfk)
 \quad , \quad \widehat{h}_{s}(\mathbf{k},\eta)=
h_{s}(k, \eta)a_{s}(\mathbf{k})+ h^{*}_{s}(k, \eta)a^{\dag}_{s}(-\mathbf{k}) \, ,
\ea
subject to the commutation relations  $ [ a_{s}(\bfk),    a_{s}^\dagger  (\bfk')] = \delta_{s s'} \delta^{(3)} (\bfk-\bfk' )$. Note that the profile of $h_s(k)$ are the same as in isotropic theory 
as given in Eq. (\ref{hs-wave}).

Based on the symmetry of our background we assume 
\ba
\bfk = k \, \big( \cos \theta \, , \sin \theta \, , 0 \big )
\ea
in which $\theta$ represents the angle between the preferred direction and the wave vector $\bfk$. 
With this convention, the polarizations $e^{+}_{ij}(\mathbf{k})$ and $e^{\times}_{ij}(\mathbf{k})$ become
\begin{align}
\label{polarization2}
e^{+}_{ij}(\mathbf{k}) =\frac{1}{\sqrt{2}} \left( \begin{array}{ccc}
 \sin^2{\theta} & -\sin{\theta}\cos{\theta} & 0 \\
  -\sin{\theta}\cos{\theta} & \cos^2{\theta} & 0 \\
  0 & 0 & -1 \\
\end{array} \right) ~~~,~~~
e^{\times}_{ij}(\mathbf{k}) = \frac{i}{\sqrt{2}}\left( \begin{array}{ccc}
  0 & 0 & -\sin{\theta} \\
  0 & 0 & \cos{\theta} \\
  -\sin{\theta} & \cos{\theta} & 0 \\
\end{array} \right) ~.
\end{align}
Using Eq. (\ref{tensor}) and Eq. (\ref{polarization2}), the components of the  tensor field operator becomes 
\begin{align}
\label{polarization3}
\widehat{h}_{ij}(\mathbf{k}) =\frac{1}{\sqrt{2}} \left( \begin{array}{ccc}
 \widehat{h}_{+}\sin^2{\theta} & -\widehat{h}_{+}\sin{\theta}\cos{\theta} & -i\widehat{h}_{\times}\sin{\theta} \\
  -\widehat{h}_{+}\sin{\theta}\cos{\theta} & \widehat{h}_{+}\cos^2{\theta} & i\widehat{h}_{\times}\cos{\theta} \\
  -i\widehat{h}_{\times}\sin{\theta} &  i\widehat{h}_{\times}\cos{\theta}& -\widehat{h}_{+} \\
\end{array} \right) \, .
\end{align}
We will use this expression later on when  calculating the correlations involving the tensor modes and the curvature perturbations.

The power spectrum of the tensor perturbations is
\ba
\langle   \widehat{h}_{ij}(\mathbf{k_1})  \widehat{h}_{ij}(\mathbf{k_2}) \rangle
= (2 \pi)^3 \delta^{(3)} (\bfk_1 + \bfk_2) P_h(k_1) \quad , \quad
\calP_h \equiv   \frac{ k_1^3  }{ 2 \pi^2}  P_h(k_1)
\ea
In the absence of anisotropy the power spectrum is given by Eq. (\ref{calPh0}).

Our aim is to calculate the corrections into the quadratic Lagrangians to read off the interaction Hamiltonian. As studied in the previous Section, the building block of the solid Lagrangian is the
symmetric matrix $B^{I J}$ in which the variables $X, Y$ and $Z$ are made of. Therefore, we have to perturb $B^{IJ}$ to second order in terms of $h_{ij}$ and $\pi_L$ perturbations.
As discussed before we do not consider  the perturbation in the transverse mode $\pi_T$. 
The reason is that the transverse excitations $\pi_T$ are not important during inflation or 
after inflation ends. As studied in  \cite{Endlich:2012pz} the wave function of $\pi_T$ is suppressed compared  to the wave function of $\zeta$ by the factor $\epsilon $. Therefore, the contribution of 
$\pi_T$ in the following interaction Hamiltonian is suppressed. Second, after inflation ends the universe become isotropic and there is no support for vector perturbations. Therefore, the vector perturbations $\pi_T$ becomes irrelevant at the time of CMB last scattering. 

We have to calculate the linear and second order corrections in $B^{IJ}$. The linear corrections in
$B^{IJ}$ are 
\begin{equation}
B_1^{IJ}=a_J^{-2}\partial _J \pi ^I +a_I^{-2}\partial _I \pi ^J-(a_Ia_J)^{-1}h _{IJ},
\end{equation}  
while the second order corrections in $B^{IJ}$ are
\begin{equation}
B_2^{IJ}=\dot{\pi}^I\dot{\pi}^J+(a_Ia_J)^{-1}h _{IK}h _{KJ}-(a_Ka_J)^{-1}h _{KJ}\partial _K \pi ^J-(a_Ka_I)^{-1}h _{IK}\partial _K \pi ^J+a_K^{-2}\partial _K \pi ^I \partial _K \pi ^J.
\end{equation}
Note that we use the convention that the repeated dummy indices (such as the index $K$ above) are
summed over while the free external indices (such as $I$ and $J$) are not summed over. 
As we shall see below,  we only need to calculate the corrections in Hamiltonian to linear order in $\sigma$ because the contributions from the terms quadratic in $\sigma^2$ in Lagrangians are
suppressed. 

Having calculated $B^{IJ}$ to second order in perturbations, we can calculate the corrections in quadratic Lagrangians. The corrections in matter Lagrangian to second order in perturbations are
\ba
\delta F &=&  \frac{ F_{XX}}{2}  (\delta_1 X)^2 
+ \frac{F_{YY}}{2}   (\delta_1 Y)^2 + \frac{ F_{ZZ}}{2}  (\delta_1 Z)^2   +  F_{XY}  \delta_1 X \delta_1 Y  +  F_{XZ}  \delta_1 X \delta_1 Z +  F_{YZ}  \delta_1 Y \delta_1 Z  
\nonumber\\
&&+ F_X \delta_2 X + F_Y \delta_2 Y + F_Z \delta_2 Z 
\ea 
in which $\delta_1 X$ and $\delta_2X$ respectively show the first order and the second order corrections in  $X$  with similar definitions for $ \delta Y$ and $\delta Z$.  As we can see from the above equation the  general form of the interaction Hamiltonian and the follow up analysis are very complicated functions of the derivatives of $F$ with respect to $X, Y$ and $Z$  and the corresponding changes in $\delta X, \delta Y$ and $\delta Z$. In order to get insights into the form of induced anisotropies, here we present the analysis for two 
important limits of the solid inflation. The first limit is the natural limit of solid as studied in \cite{Endlich:2012pz} in which $| F_Y| \sim |F_Z| \sim |F|$ subject to the condition Eq. (\ref{FYZ-cond}).  The second limit is the opposite of the above limit in which $F= F(X)$ so 
$F_Y= F_Z=0$.  We present the results in both of these limits which show similar patterns. 

\subsection{Solid with $ F_Y, F_Z \sim F$}
\label{first-limit}

The limit $ |F_Y|, |F_Z| \sim |F|$,  subject to the condition Eq. (\ref{FYZ-cond}), 
is considered as the natural limit of solid inflation in  \cite{Endlich:2012pz}.  In this limit we have 
$F_Y = - F_Z + O( \epsilon)$,   $X F_X = \epsilon F,  X^2 F_{XX} \simeq -\epsilon F  $  and $F_{XY} \sim F_{X Z} \sim \epsilon F$ so 
we can safely neglect the terms in $\delta F$ containing derivative of $F$ with respect to $X$. In addition 
$F_{YY}\simeq   F_{ZZ} \simeq - F_{YZ}$ so to leading order in slow-roll parameters 
the corrections in Lagrangian are
\ba
\label{delta F-YZ0}
\delta F \simeq F_Y \big(  \delta_2 Y - \delta_2 Z  \big) + \frac{F_{YY}}{2} \big(  \delta_1 Y - \delta_1Z  \big)^2 \, .
\ea
One can check that $\delta_1 Y- \delta_1 Z = {\cal O} (\sigma^2)$ so the contribution from the second term in Eq. (\ref{delta F-YZ0})  containing $F_{YY}$ is at the order of $ \sigma^4$ which are
quite negligible in the limit $\sigma \ll 1$. 

After a long calculation  $ \delta_2 Y - \delta_2 Z$  to leading order in $\sigma$  is obtained to be  
\ba
\delta_2 Y-\delta_2 Z &=&-\frac{2}{9}\sigma \left( 2B^{xi}_1B_1^{xi}-B_1^{yi}B_1^{yi}-B_1^{zi}B_1^{zi}\right)+\frac{4}{27}\sigma \delta_1X \left(2B_1^{xx}-B_1^{yy}-B_1^{zz}\right) \nonumber \\
&+&\frac{16}{27} \delta_1X \sigma ^2 \left(4B_1^{xx}+B_1^{yy}+B_1^{zz}\right)+\frac{8}{9}\sigma ^2 \delta_2 X-\frac{32}{27}\sigma ^2 (\delta_1X)^2
+ {  \frac{4}{9}} \sigma ^2 B_1^{ij}B_1^{ij}
  \nonumber \\
&-&\frac{4}{9}\sigma ^2 \left(4B_2^{xx}+B_2^{yy}+B_2^{zz}\right) -\frac{2}{9}\sigma ^2\left( 4 B_1^{xi}B_1^{xi}+B_1^{yi}B_1^{yi}+B_1^{zi}B_1^{zi}\right). \label{dY-dZ}
\ea
As mentioned before, we need the terms linear in $\sigma$ to calculate the leading order anisotropy, 
but we kept the terms quadratic in $\sigma^2$ for the future references in our discussions.  
  
The quadratic Lagrangians density  responsible for anisotropies  to  $O(\sigma)$ are 
\ba
\label{LSS}
\delta{\cal L}_{\zeta \zeta }
&=& -\frac{8}{27} \sigma F_Y  \nabla ^2 \pi _L \left(2\partial _x^2 \pi _L - \partial _y^2 \pi _L\right) \, ,
\\
\label{LTT}
\delta{\cal L}_{h h}&=&-\frac{2}{9}\sigma  {F_Y } \left( 2h _{xx}^2+h _{xy}^2+h _{xz}^2-
2h _{yz}^2-h _{zz}^2-h _{yy}^2\right)  \, ,
\\
\label{LST}
{\cal L}_{\zeta h} &=&{-\frac{8}{9}\sigma  F_Y  \nabla^2 \pi_L h_{xx} } \, .
\ea
Note that $\delta {\cal L}_{\zeta \zeta }$ and $\delta {\cal L}_{h h }$ respectively represents the corrections in the quadratic scalar  and tensor Lagrangians while $\delta{\cal L}_{\zeta h}$ represents the Lagrangian mixing the scalar and the tensor at the quadratic level. In particular, note that the scalar-tensor cross-correlation exits only
in the anisotropic background as sourced by ${\cal L}_{\zeta h}$.

It is convenient to write down the  interaction Lagrangian density in Fourier space. Using 
the relation $\zeta = -k^2 \pi_L/3$, and the explicit form of the tensor components as given in 
Eq. (\ref{polarization3}) we have 
\ba
\label{delta-Lss}
\delta \cal{L}_{ \zeta \zeta } &=& \frac{8 \sigma}{3} F_Y (1- 3 \cos^2 \theta) | \zeta|^2 \, , \\
\label{delta-LTT}
\delta {\cal L}_{ hh } &=&   -\frac{ \sigma}{9} F_Y  (1- 3 \cos^2 \theta) \left(  |h_+|^2 + | h_\times|^2   \right) \, , \\
\label{L-ST}
{\cal L}_{\zeta h} &=&  -\frac{4 \sigma F_Y}{3 \sqrt2} \sin^2 \theta \left( \zeta h_+^* + c.c. \right) \, .
\ea

Having calculated the leading interaction Lagrangians induced from the anisotropies we are ready to calculate the anisotropic corrections in scalar power spectrum $\delta P_\zeta$, the anisotropic corrections in tensor power spectrum $\delta P_h$ and the scalar-tensor cross-correlation $P_{\zeta h}$ using the standard  
 in-in formalism  \cite{ Weinberg:2005vy, Chen:2009zp, Chen:2010xka, Wang:2013zva}. 
For this purpose we have to use the interaction Hamiltonian $H_I$. However, for our model with no kinetic coupling between the fields one can easily check that $H_I=-L$ in which $L$ are the interaction Lagrangians given in Eqs. (\ref{delta-Lss}),  (\ref{delta-LTT}) and (\ref{L-ST}).

\subsubsection{Anisotropy in Curvature Perturbations Power Spectrum}
\label{zeta-power}

\begin{figure}[t]
  \centering
  \includegraphics[width=0.5\textwidth]{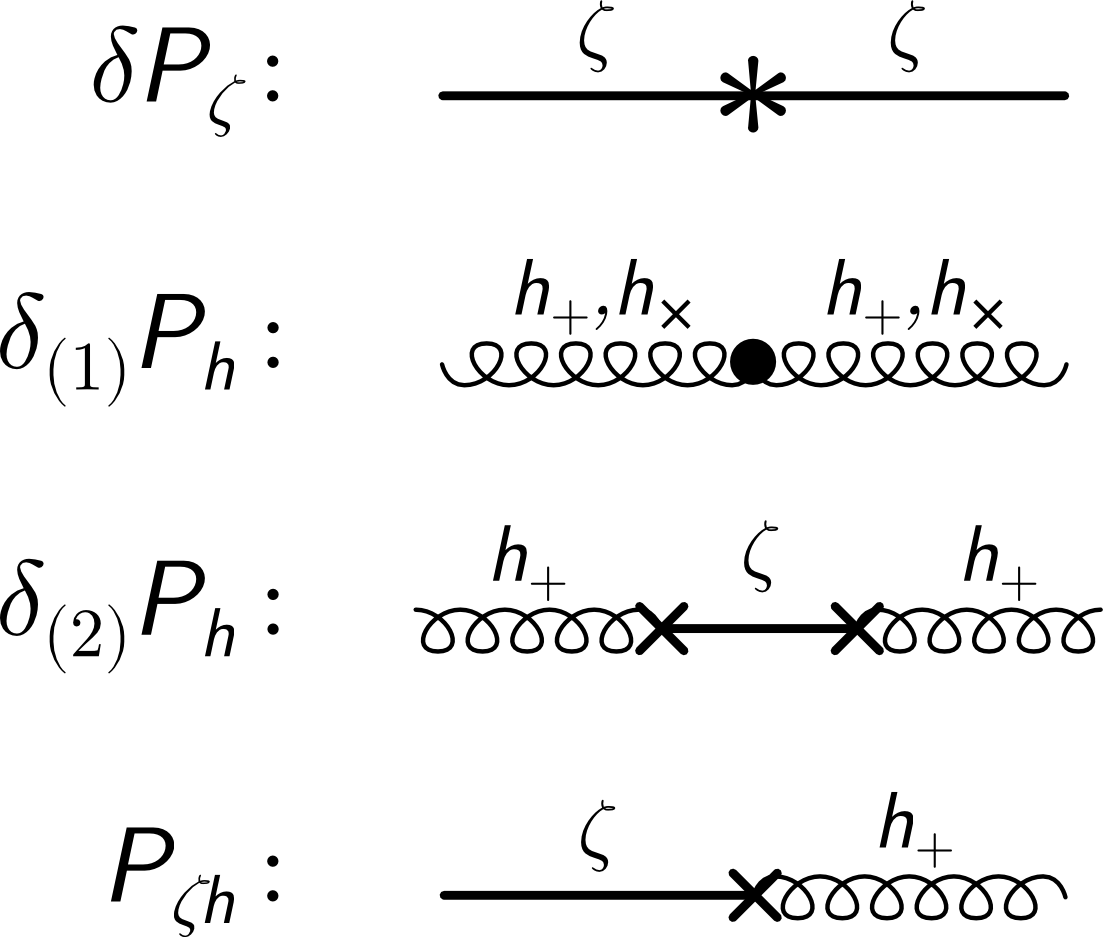}\vspace{0.5cm}
  \caption{\label{fig:feynman} Here we present the Fyenman diagrams. The top diagram 
  corresponds to $\delta {\cal L}_{\zeta \zeta}$ which can be interpreted as the change in
  $\zeta$ effective mass. The second diagram corresponds to $\delta {\cal L}_{h h}$
  which also is in the form of a mass insertion.  The third diagram represents the corrections
  in $h_+$ propagator from the insertion of two exchange vertices. The last diagram represents the
  scalar-tensor cross-correlation. Note that the three couplings are  indicated by
  $*, \bullet$ and $\times$. }
\end{figure}

First we calculate the anisotropy in curvature perturbation power spectrum 
$\delta P_\zeta$.  As discussed before the free theory corresponds to the isotropic solid inflation with the wave function of $\zeta$ given in Eq. (\ref{zeta}).  The leading contributions in $\delta P_\zeta$
come from $L_{\zeta \zeta}$ which is linear in $\sigma$.  The corresponding Feynman diagram is shown in Fig.~\ref{fig:feynman}. Intuitively speaking, this diagram corresponds to corrections in scalar perturbations effective mass.  Using the standard in-in formalism, we have
\ba
\delta P_\zeta  &=& -i \int_{\eta_0}^{\eta_e} d\eta a(\eta)^4 { \bigg\langle}  {\bigg[} \delta L_{SS} ~ , ~ \zeta(\eta_e) \zeta(\eta_e)^* \bigg] { \bigg \rangle } \nonumber\\
&=& \frac{32 \sigma F_Y}{3 H^4}(1 - 3 \cos^2 \theta)  \int_{\eta_0}^{\eta_e} \frac{d \eta}{\eta^4}
\im \bigg[ \big\langle  \zeta(\eta) \zeta(\eta_e)^*  \big\rangle  \big\langle  \zeta(\eta) \zeta(\eta_e)^* 
 \big\rangle  \bigg]  
\ea
in which $\eta_0$ represents the initial time of inflation when the modes of interest were deep 
inside the (sound) horizon  $ k c_L \eta \ll -1 $ and $\eta_e$ indicates the time of end of inflation in which $k \eta_e \rightarrow 0^-$. 
Using $\eta(\eta_e) \simeq {i C \sqrt2}/{c_L^2 \sqrt\pi}$ the above integral is translated into the following dimensionless integral
\ba
 \im \int_{-\infty}^{0} \frac{d x}{x^4}  \left[ e^{-2 i x} (x^2 - 3 i x -3)^2  \right]  = \frac{5}{2}
\ea
in which we have used the replacement $x \rightarrow x+ i  x \delta$ for the contour of the integral
with $\delta \rightarrow 0^+$.  Putting all together we obtain 
\ba
\label{delta-P-zeta}
\delta P_\zeta = \frac{5 \sigma F_Y}{27 M_P^4 \epsilon^2 k^3 c_L^7} (1 - 3 \cos^2 \theta)
\ea
Now using
\ba
P_\zeta^{(0)}  = \frac{H^2}{4 \epsilon c_L^5 M_P^2 k^3} \, ,
\ea
we can express $\delta P_\zeta $ as a fraction of $P_\zeta^{(0)}$ as follows
\ba
\label{delta-P-zeta2}
\delta P_\zeta = \frac{-20}{9}\frac{\sigma F_Y}{\epsilon F c_L^2} ( 1- 3 \cos^2 \theta )  \, 
P_\zeta^{(0)} \, .
\ea
Interestingly, this is the same result as  obtained in \cite{Endlich:2013jia} using the peak-background splitting in the three-point function $\langle h_s \zeta^2 \rangle $ treating a long perturbation 
in $h_s$ as a change of effective background for $\zeta$ perturbations  (note that  
our $\sigma$ is $-1/2$ of $\sigma$ used in \cite{Endlich:2013jia}).  
 
The quadrupole asymmetry induced in curvature power spectrum is the hallmark of Bianchi I background. Similar situation arises in models of anisotropic inflation with the $U(1)$ gauge field
with the Lagrangian $-f(\phi)^2 F_{\mu \nu} F^{\mu \nu}$ as studied in 
\cite{Watanabe:2010fh}, for related works on primordial anisotropies see \cite{Pereira:2007yy}.


We define the amplitude of the quadrupole asymmetry $g_*^{\zeta}$  via
\ba
P_\zeta = P_\zeta^{(0)} \big( 1+ g_*^{\zeta} (\widehat \bfk \cdot \widehat{\mathrm{\bf n}})^2 
\big) \, ,
\ea
in which $ P_\zeta^{(0)}$ is the isotropic curvature perturbations power spectrum and $\widehat{\mathrm{\bf n}}$ represents the preferred direction in the sky (the $x$-axis in our example).  Note that historically $g_*^\zeta$
is defined as $g_*$ in models of anisotropic inflation with the assumptions that the tensor perturbations are negligible. However, in the presence of GW one has to be careful as what one means
by $g_*$. We define $g_*$ as the amplitude of quadrupole anisotropies in total CMB temperature map which contains both scalar and tensor perturbations and their cross-correlations.  Therefore,  
$g_*$ contains not only the contributions from $\delta P_\zeta$ but also from the anisotropy in tensor
power spectrum $\delta P_h$ and the scalar-tensor cross-correlation $P_{\zeta h}$. This point was 
discussed in \cite{Chen:2014eua}.  Therefore, in order to prevent confusion, we denote the amplitude of quadrupole asymmetry in $P_\zeta$ by $g_*^\zeta$ with similar definition for $g_*^h$ and $g_*^{\zeta h}$.  

With these discussions, the anisotropic parameter $g_*^{\zeta}$ therefore is
\ba
g_*^\zeta = \frac{20}{3} \frac{\sigma F_Y}{\epsilon c_L^2 F}
\ea
We see that $g_*^\zeta$ can take  either signs depending on the sign of $\sigma F_Y/F$. One interesting feature of the above formula is that to leading order $g_*^\zeta$ is independent of $N$, the  number of e-folds when the mode of interest $k$ has left the horizon till end of inflation. This should be compared with $g_*^\zeta$ obtained in anisotropic inflation in which $g_*^\zeta \propto N^2$. This indicates that to leading order, unlike models of anisotropic inflation  \cite{Bartolo:2012sd}, the scalar power spectrum does not suffer from the IR anisotropies.  We will come back to this issue when studying anisotropies in tensor perturbations. 

Note that the next corrections in $\delta P_\zeta$ and $g_*^{\zeta}$ are
at the order of $\sigma^2$ and are suppressed compared to the leading term given in Eq. (\ref{delta-P-zeta}). These sub-leading corrections have two sources. They can come either directly from the $\sigma^2$ corrections in ${\cal L}_{\zeta \zeta}$ or from the inclusion of two exchange vertices 
${\cal L}_{\zeta h }$ in the propagator of $\zeta$. 

As discussed before, solid inflation predicts large non-Gaussianities with similar squeezed limit to the local shape. This is in contrast to the celebrated Maldacena's consistency condition for single field model  \cite{Maldacena:2002vr}.
In the limit considered here with $F_Y = -F_Z$, the non-Gaussianity parameter $f_{NL}$ is calculated in \cite{Endlich:2012pz} yielding
\ba
\label{f-NL}
f_{NL} \simeq -\frac{25 F_Y}{27  \epsilon c_L^2 F}(1- 3 \cos^2 \theta) \, .
\ea
For $F_Y \sim F$ this results in large non-Gaussianities which are  directional-dependent. There are strong
upper bound on $f_{NL}$ from the Planck data, $f_{NL} = 2.7\pm 5.8$ (68 \% CL) \cite{Ade:2013ydc}. Therefore, in order to be consistent with this bound, $F_Y/F$ should be much smaller than unity. Having this said, we mention that the Planck constraint on $f_{NL}$ may not be directly applicable to the solid model in which the bispectrum has non-trivial directional-dependence. Therefore, the Planck constraint on $f_{NL}$ for this case can be used as a rough order of magnitude estimate. 

Comparing Eqs. (\ref{delta-P-zeta2}) and (\ref{f-NL}) we see the interesting result that 
\ba
\frac{\delta P_\zeta}{P_\zeta^{(0)}} = \frac{12}{5} \sigma f_{NL} \, .
\ea
In addition, neglecting the directional dependence in  $f_{NL}$, we have $g_*^\zeta \sim \sigma f_{NL}$.  This result also indicates that in order not to produce too much anisotropies the amplitude of local-like non-Gaussianities should be under control.

\subsubsection{Anisotropy in Tensor Power Spectrum}

Now we calculate anisotropies generated in tensor power spectra $\delta P_h$. 

There are two contributions in $\delta P_h$.
The first contribution is linear in $\sigma$ and comes directly from the corrections in the tensor   quadratic action $\delta {\cal L}_{h h}$. The corresponding Feynman diagram is shown in Fig.~\ref{fig:feynman}, the second diagram from top. Intuitively speaking, this diagram corresponds to change in tenor perturbation effective mass.  The second 
contribution in $\delta P_h$ is quadratic in $\sigma^2$ which comes from two exchange vertices ${\cal L}_{\zeta h}$ inside the $h_+ $ propagator. The corresponding Feynman diagram is the third diagram from top  in Fig.~\ref{fig:feynman}. As discussed in previous sub-section we have neglected the $\sigma^2$ corrections in $\delta P_\zeta$. However, one can not simply neglect the $\sigma^2$ corrections in $\delta P_h$. The reason is that the wave function of $\zeta$ is enhanced by a factor $\epsilon^{-1/2}$ compared to  the wave function of $h_s$ so the relative ratio of the two contributions in $\delta P_h$ is at the order $(\sigma/\epsilon) F_Y/F$ which may not be negligible (actually we shall see that this ratio is more enhanced by additional factor $N/c_L^5$ in which 
$N$ is the total number of e-folds).  Having this said, we do not have to calculate $\sigma^2$ corrections from the higher order corrections in ${\cal L}_{hh}$ as they are suppressed compared to the linear term by additional  factor of $\sigma$.

Let us start with the anisotropic corrections  linear in $\sigma$, denoted by $\delta_{(1)} P_{h}$.
To linear order in $\sigma$ both polarizations $h_+$ and $h_\times$ appear symmetrically in $\delta L_{hh}$ in Eq. (\ref{delta-LTT})
so  it is enough to calculate the change in power spectrum of either 
polarizations and multiply it bay factor 2. Choosing $h_+$  we have 
\ba
\delta_{(1)} P_{h_+}  &=& -i \int_{\eta_0}^{\eta_e} d\eta a(\eta)^4 { \bigg\langle}  {\bigg[} \delta L_{hh} ~ , ~ h_+(\eta_e) h_+(\eta_e)^* \bigg] { \bigg \rangle } \nonumber\\
&=& -\frac{4 \sigma F_Y}{9 H^4}(1 - 3 \cos^2 \theta)  \int_{\eta_0}^{\eta_e} \frac{d \eta}{\eta^4}
\im \bigg[ \big\langle  h_+(\eta) h_+(\eta_e)^*  \big\rangle  \big\langle  h_+(\eta) h_+(\eta_e)^* 
 \big\rangle  \bigg]  
\ea
Using the form of wave function for tensor perturbations as given in Eq. (\ref{hs-wave}) the above integral is cast into the following form in terms of the dimensionless variables
$x= k \eta$
\ba
\label{int2}
\int_\infty^0 \frac{dx}{x^4} e^{-2 i x} (x -i)^2 \simeq -\frac{2 N}{3}
\ea
in which $N= - \ln (- k \eta_e)$ is the number of e-folds when the mode $k$ leaves the horizon till
end of inflation. Note that we have neglected order one corrections in Eq. (\ref{int2}) which is justified
for $N \gg 1$. 

The anisotropy in tensor power spectrum at linear order in $\sigma$ is obtained to be
\ba
\delta_{(1)} P_h =  2 \delta_{(1)} P_{h_+} =  \frac{64 H^2}{9 M_P^2 k^3} \frac{\sigma N F_Y}{F} (1- 3 \cos^2 \theta) \, .
\ea

Now we calculate the $\sigma^2$ corrections in tensor power spectrum denoted by   
$\delta_{(2)} P_{h}$. This comes from the Feynman diagram shown in Fig.~\ref{fig:feynman}  yielding
\ba
\delta_{(2)} P_h &=& -\int_{\eta_0}^{\eta_e} d \eta_1 d \eta_2 \,  a(\eta_1)^4 \, a(\eta_2)^4
\bigg \langle \bigg[ L_{\zeta h_+} \,  ,  \,  \bigg[  L_{\zeta h_+} \, , \,  h_+(\eta_e) h_+(\eta_e) \bigg] \, \bigg] \bigg \rangle \\
&=& 32 \left(  \frac{4 \sigma }{3\sqrt2 H^4}  F_Y \sin^2 \theta   \right)^2 
\int_{\eta_0}^{\eta_e} \frac{d \eta_1}{\eta_1^4} \frac{d \eta_2}{\eta_2^4} 
\im \bigg[  h(\eta_1) h(\eta_e)^* \bigg] \im \bigg[  \zeta(\eta_2 ) \zeta(\eta_1)^* h(\eta_2) h(\eta_e)^*
\bigg] \, , \nonumber
\ea
in which the factor $32$ comes from the symmetry and permutations considerations. The above integral can be taken after performing the appropriate contour rotation for $\eta \rightarrow -\infty$ and taking $\eta_e \rightarrow 0$. One important point to observe is that the arguments of Hankel functions for $h(\eta)$
and $\zeta(\eta)$ are different by factor of $c_L$. This plays important role. For example, if one naively takes both arguments of the Hankel functions to be equal, corresponding to simply setting $c_L=1$,   then the final result will be off by factor of $c_L^{-6} \ge 27 $.

Taking the integral we get 
\ba
\label{delta2-h}
\delta_{(2)} P_h = \frac{128 H^2}{9M_P^2 k^3} \frac{\sigma^2 N^2 F_Y^2}{\epsilon c_L^5 F^2} 
\sin ^4 \theta  \, .
\ea
There are few interesting observations here. First, we see that the anisotropy in $\delta_{(2)} P_h $
has a different shape than $\delta_{(1)} P_h $  or $\delta P_\zeta$, it has $\sin^4 \theta$ instead
of the usual $\cos^2 \theta $ corrections. Second, the amplitude is proportional to $N^2 (F_Y/F)^2$.
The conclusion that the amplitude is quadratic in  $F_Y/F$ is expected since we have inserted two exchange vertices of ${\cal L}_{\zeta h}$. The factor $N^2$ comes from the fact that we have two nested integral involving the tensor mode $h_+$. Compare this with the integral in 
$\delta_{(1)} P_h $ which involves a single integral containing $h_+$ which also yields the factor N.

Combining the two contributions in $\delta P_h$ yields
\ba
\label{Ph-total}
\delta P_h = \delta_{(1)} P_h +  \delta_{(2)} P_h
&\simeq&   \frac{256 N \sigma F_Y}{9 F}  \left[ \epsilon c_L^5 (1 - 3 \cos^2 \theta)
 {+} 2  \sigma N \frac{F_Y}{F}  \sin ^{{4}} \theta \right] P_\zeta^{(0)} \nonumber\\
 &=& \left( \frac{64}{15} N g_*^\zeta c_L^4 \right)  \epsilon^2    \left[  c_L^3 (1 - 3 \cos^2 \theta)
 {+}\frac{3}{10}   N g_*^\zeta  \sin ^{{4}} \theta \right] P_\zeta^{(0)}
\ea
It is instructive to compare $\delta_{(2)} P_h$ with $\delta_{(1)} P_h$. Considering only the amplitude we have 
\ba
\label{Ph-ratio}
\frac{\delta_{(2)} P_h}{\delta_{(1)} P_h} \simeq  \frac{2  \sigma N F_Y}{\epsilon c_L^5 F} \simeq 
\frac{3 N g_*^\zeta }{10 c_L^3}
\ea
Therefore, depending on the observational bound on $g_*^\zeta$, this ratio can be bigger than one.
For example, suppose we take $|g_*^\zeta| = 0.1$. Then the above ratio is at the order $2/c_L^3 \gtrsim 10$.

\subsubsection{The Scalar-Tensor Cross-Correlation}
\label{zeta-j-power}

Now we calculate the scalar-tensor cross-correlation $\langle \zeta h \rangle$  spectrum, $P_{\zeta h}$. The leading contribution in $P_{\zeta h}$ comes from the last Feynman diagram 
shown in Fig.~\ref{fig:feynman} 
\ba
P_{\zeta h}  &=& -i \int_{\eta_0}^{\eta_e} d\eta \, a(\eta)^4 { \bigg\langle}  {\bigg[} \delta L_{\zeta h} , h_+(\eta_e) \zeta(\eta_e)^* \bigg] { \bigg \rangle } \nonumber\\
&=&-\frac{16 k^2 \sigma}{3 \sqrt2} F_Y \sin^2 \theta  \int_{\eta_0}^{\eta_e} d\eta \, a(\eta)^4 \, 
\im \bigg[   \bigg\langle \zeta(\eta) \zeta(\eta_e)^* \bigg \rangle \bigg \langle h_+ (\eta) h(\eta_e)^* \bigg \rangle  \bigg] \nonumber\\
&=& -\frac{8}{9 \sqrt2} \frac{N \sigma F_Y}{\epsilon c_L^5 k^3 M_P^4 } \, .
\ea
As a result
\ba
\label{P-zeta-h}
P_{\zeta h} &=&  \frac{32}{3 \sqrt2} \frac{N \sigma F_Y}{F} \sin^2 \theta   P_\zeta^{(0)} \nonumber\\
&=&  \left( \frac{8}{5 \sqrt2} N g_*^\zeta c_L^2    \right) \epsilon \sin^2 \theta P_\zeta^{(0)} \, .
\ea
Depending on the sign of $g_*^\zeta$ the cross-correlation $\langle \zeta h \rangle$ can be either a correlation (positive sign) or an anti-correlation (negative sign). Note that the amplitude of $P_{\zeta h} $ is bigger than the amplitude of $\delta P_h$ by a factor $1/\epsilon$. This means that in the temperature anisotropy power spectrum  the tensor-scalar cross-correlation is more significant than the statistical 
anisotropies induced from the  tensor power spectrum.  

\subsubsection{Imprints on TT Correlations  }

So far we have calculated $\delta P_\zeta, \delta P_h$ and $\delta P_{\zeta h}$.
To get observable effects in the $TT$ or $TB$ correlations one has to allow for
large enough value of $g_*^\zeta$. This calls for a careful analysis of CMB constraint on $g_*^\zeta$. In \cite{Kim:2013gka} (see also \cite{Ramazanov:2013wea}) this was 
performed for the Planck data with the assumption of no tensor perturbations so 
$g_* = g_*^\zeta$,  yielding the upper bound $|g_*| \lesssim 10^{-2}$. However, in the presence of tensor mode this analysis has to be redone. As argued in \cite{Chen:2014eua} the effective value of $g_*$ is different than $g_*^\zeta$. The physics behind this difference is that the tensor anisotropies contribute on the 
low-$\ell$ multipoles while their contributions on higher $\ell$ dies off rapidly because of their decaying transfer function. As a result, this induces a non-trivial scale-dependence in $TT$ anisotropies so the results obtained in \cite{Kim:2013gka} can not be used directly. Therefore, it will be interesting to consider the predictions of solid inflation on $g_*^\zeta$ and then see how significant the results of tensor modes in $TT$, $TB$ and $BB$ correlations are. 

With these discussions in mind, one may ask how significant the contribution of the anisotropic tensor modes in $TT$ correlation is. In a rough estimation one may assume the CMB temperature fluctuations  has contributions from the scalar and tensor parts as 
$\delta T = \delta T^\zeta + \delta T^h$. Neglecting the non-trivial contributions of transfer function 
this yields 
\ba
\left \langle\,  (\delta T )^2 \, \right  \rangle     \sim  \left \langle \, (\delta T^\zeta )^2 \,  \right \rangle \left(1+ g_*^\zeta  \cos^2 \theta 
+ 2 \frac{P_{\zeta h}}{P_\zeta^{(0)}} +  \frac{\delta P_{ h}}{P_\zeta^{(0)}}  \right) \, .
\ea
Therefore, in order to estimate the effects of tensor mode anisotropies in the TT correlation we have to compare the ratios  $\frac{P_{\zeta h}}{P_\zeta^{(0)}}$ and  $\frac{\delta P_{ h}}{P_\zeta^{(0)}}$
with $g_*^\zeta$. Using Eq. (\ref{P-zeta-h}) we find  
$\frac{P_{\zeta h}}{ g_*^\zeta P_\zeta^{(0)}} \sim N \epsilon$ while  from
Eq. (\ref{Ph-total}) we have  $\frac{\delta P_{ h}}{g_*^\zeta P_\zeta^{(0)}} \sim (N \epsilon)^2 g_*^\zeta$.  Assuming $N \simeq 60$ and $N\epsilon \sim 1$, we see that the contribution of $P_{\zeta h}$ in $TT$ correlation can be  important while the contributions of  $ \delta P_h$ seems too small to be important.  We will study the imprints of $\delta P_\zeta$, $\delta P_h$ and $\delta P_{\zeta h}$ on CMB spectra in details in next Section.

As discussed above the contributions of  tensor perturbations in CMB correlations decay on large $\ell$. This property was employed in \cite{Chen:2014eua} to speculate  that the tensor perturbations may be behind various anomalies observed in the TT correlations. It is argued that  the  shortage of power in low multipoles and  the  dipole asymmetry for  $\ell < 64$ in CMB temperature map may be related to the B-mode polarization as
detected by the BICEP2 observation. In particular, it is argued in \cite{Chluba:2014uba} that 
the  hemispherical asymmetry in tensor modes which are generated from the long mode modulations  \cite{Dai:2013kfa, Abolhasani:2013vaa} may alleviate   the  apparent tension between the BICEP2 and  Planck observations.

\subsubsection{Comparison to Models of Anisotropic Inflation }

It is instructive to compare the results obtained here with the models of anisotropic inflation studied 
in \cite{Ohashi:2013qba} and \cite{Chen:2014eua}. The model of anisotropic inflation studied in
\cite{Ohashi:2013qba} corresponds to a real inflaton field coupled to the gauge field via the Lagrangian $- f(\phi)^2 F_{\mu \nu} F^{\mu \nu}/4$. To get an attractor solution with a sub-dominant 
contribution of the gauge field energy density to the total energy density we require $f(\phi) \propto a(t)^{-2}$. The corrections  in tensor power spectrum is related to $g_*^\zeta$ via
 $\delta P_{h}/{P_h^{(0)}} \simeq \epsilon g_*^\zeta/4$.
With $g_*^\zeta  <1$, the ratio  $\delta P_{h}/{P_h^{(0)}}$ is very small so the anisotropic effects in tensor power spectrum is perhaps beyond detection. 

Now consider the model studied in \cite{Chen:2014eua} in which the inflaton is a complex scalar field charged under the $U(1)$ gauge field via the gauge coupling $\mathrm{\bf e}$. Interestingly, there is 
not strong constraint on $\mathrm{\bf e}$ from the background or from  $g_*^\zeta$. One may take $ \mathrm{\bf e} \sim 10^{-3}$ consistent with observational bound on $g_*^\zeta$. The anisotropy correction in tensor power spectrum is calculated to be
$\delta P_{h}/{P_h^{(0)}} \simeq g_*^\zeta/4 \epsilon$. With $g_*^\zeta \sim \epsilon$ this ratio can easily reach order unity so the perturbative approach assuming that the anisotropic correction is small 
breaks down.   Therefore, it was concluded in \cite{Chen:2014eua} that the tensor mode is 
a sensitive probe of the gauge coupling $\mathrm{\bf e}$. The anisotropic effects in tensor modes may be detected in  $TT, TB $ and $BB$ correlations in the upcoming Planck polarization maps. 

Now let us look at the ratio  $\delta P_{h}/{P_h^{(0)}}$ in our model. Using Eq. (\ref{Ph-total})
and taking $c_L^2 \sim 1/3$ for simplicity
we get  
\ba
\label{Ph-ratio}
\frac{\delta P_h}{P_h^{(0)}} \simeq  \bigg(\frac{3 \sqrt3}{10}  N^2 g_*^\zeta \bigg) \left(\frac{g_*^\zeta \epsilon}{4} \right) \, .
\ea
Comparing to the results of \cite{Ohashi:2013qba} we have the additional factor  $\big(\frac{3 \sqrt3}{10}  N^2 g_*^\zeta \big)$. Taking $N=60$ and $g_*^\zeta =1/10$ we get 
$\delta P_{h}/{P_h^{(0)}} \simeq \epsilon$. This is about one or two orders of magnitude larger than the results in  \cite{Ohashi:2013qba} for models of anisotropic inflation with a real inflaton field. However, the ratio   $\delta P_{h}/{P_h^{(0)}} $ for solid inflation 
is typically very small so our perturbative treatment 
is consistent.

Models of anisotropic inflation based on $U(1)$ gauge fields suffer from the IR anisotropies \cite{Bartolo:2012sd}. This effect corresponds to gauge fields fluctuations which has left the horizon
in the past inflationary history and become classical afterwards. The accumulated IR anisotropies  add up making the background more and more anisotropic. As studied in models of anisotropic inflation \cite{Ohashi:2013qba, Chen:2014eua} the IR anisotropies in $\delta P_\zeta, \delta P_h$ and $\delta P_{\zeta h}$ grows like $N^2$.  Therefore,  if inflation lasts indefinitely long in the past, the IR anisotropies can dominate over the classical background making the universe completely anisotropic. This invalidate our starting assumption in taking anisotropies to be sub-leading corresponding to $g_*, g_*^\zeta \ll 1$. 

Now let us look at  IR anisotropies in anisotropic solid inflation.  From Eq. (\ref{delta-P-zeta2}) we see that to lading order $\delta P_\zeta$ is independent of $N$. On the other hand from 
Eqs. (\ref{Ph-total}) and (\ref{P-zeta-h}) we see that  the leading terms in 
$\delta P_h$ and $\delta P_{\zeta h}$ are proportional to $N^2 $ and $N$ respectively. Therefore the 
IR anisotropies are more pronounced in tensor perturbations than in scalar perturbations.
At first look, one may conclude that if inflation continues very long in the past, then the anisotropies 
in tensor perturbation becomes very large so our perturbative approach breaks down. However, the situation is somewhat tricky in solid background. From Eq. (\ref{sigma-t}) we see that if inflation extends  for a long period, then $\sigma$ decays like $e^{-(4/3) c_T^2 N \epsilon }$ in which we have neglected  the running of $\epsilon$ and $c_T$. Therefore, there will be a balance between the level of IR anisotropies and the duration of inflation. Specifically, the measure of IR anisotropies in tensor perturbations is $N \sigma \sim N e^{-(4/3) c_T^2 N \epsilon}$. Therefore, as just mentioned, there is a competition between $N$ and the level of anisotropy. For example taking $(4/3) c_T^2\epsilon  \sim 0.05$ then $N \sigma$ reaches a maximum value for $N  \sim 20 $ and then decays mildly. In particular, if one waits long enough, then the IR anisotropies are completely washed out and we 
reach the FRW regime as the attractor solution of  Eq. (\ref{sigma-t}). The time-scale for IR anisotropies to reach the maximum is $N _{max} \sim 1/\sqrt{\epsilon } c_T$.

\subsection{$F(X)$ Model of Solid Inflation}
\label{FX-model}

In the previous sub-sections we have studied the limit $F_Y \sim F_Z \sim F$ as the natural limit of 
solid inflation. The other  limit of interests correspond to the extreme situation in which
$F= F(X)$ and $F_Y=F_Z=0$. As studied in \cite{Arroja:2010wy, Chen:2013kta},  this may be interpreted as the  fluid description of solid inflation.

The analysis in this limit is simplified since we only deal with $\delta F = F_X \delta_2 X + F_{XX} (\delta_1 X)^2/2$. Now calculating the linear and quadratic corrections in $X$, 
$\delta_1 X$ and $\delta_2 X$, we have  
\ba
\delta_1 X = e^{-2 \alpha} \sigma \bigg[  8  \partial^2_x \pi_L - 4   \partial^2_y \pi_L
+ 6  h_{xx} + 2 \nabla^2 \pi_L \bigg]
\ea
and
\ba
\delta_2 X = 2 \sigma e^{-2 \alpha} \bigg[   2h _{xx}^2+h _{xy}^2+h _{xz}^2-2h _{yz}^2-h _{zz}^2-h _{yy}^2 \bigg]  + 2 \sigma e^{-2 \alpha}
\nabla^2 \pi_L \left(  2  \partial^2_x \pi_L -    \partial^2_y \pi_L
\right)
\ea
As a result, the interaction Lagrangians are
\ba
\delta {\cal L}_{\zeta \zeta} &=& {  {\frac{-2\sigma}{3}} } F_X e^{-2 \alpha} \nabla^2 \pi_L \left(  2  \partial^2_x \pi_L -    \partial^2_y \pi_L \right) \\
\delta{ \cal L}_{h h} &=& 2 \sigma F_X e^{-2\alpha}  \bigg[   2h _{xx}^2+h _{xy}^2+h _{xz}^2-2h _{yz}^2-h _{zz}^2-h _{yy}^2 \bigg] \\
{\cal L}_{\zeta h} &=& { { 4 \sigma  F_X } } e^{-2 \alpha} h_{x x} \nabla^2 \pi_L 
\ea
Going to Fourier space we get 
\ba
\delta \mathcal{L}_{\zeta \zeta} &=&-6 \epsilon \sigma M_p^2 H^2 
{ { |\zeta |^2 }}   \left(1- 3\cos ^2 \theta \right),
\\
\delta \mathcal{L}_{\zeta h} &=&-3 \sqrt{2}\epsilon M_p^2 H^2 \sigma \sin ^2 \theta  
\left(  {   \zeta} \,  h_{+}^{*}+c.c\right),
\\
\delta \mathcal{L}_{hh} &=&-\epsilon \sigma M_p^2 H^2 (1-3\cos ^2 \theta) \left( |h_{+}|^2+|h_{\times}|^2\right).
\ea

Now we can calculate the anisotropic corrections in power spectra and in 
scalar-tensor cross-correlation. Happily, the form of the interaction Lagrangians are the same as in Eqs. (\ref{delta-Lss}),  (\ref{delta-LTT}) and (\ref{L-ST}) so we do not need to perform the in-in integrals again and we only  have to take into account the difference in numerical factors. Also note that in the limit $F=F(X)$ we have $c_L^2 =\frac{1}{3} + O(\epsilon, \eta)$.

For the anisotropic corrections in curvature perturbation power spectrum $\delta P_\zeta$  we have
\begin{equation}\label{eq:delta-P-zeta-FX}
\delta P_{\zeta}=     5 \sigma (1- 3 \cos^2 \theta) P_\zeta^{(0)}
\end{equation}
As a result, the anisotropic parameter $g_*^\zeta$ has the simple form 
\ba
g_*^\zeta = 15 \sigma \, .
\ea
This is an interesting result indicating that $g_*^\zeta$ is independent of the form of $F(X)$.

For the anisotropic corrections in tensor power spectra,  $\delta _{(1)} P_{h}$ and $\delta _{(2)} P_{h}$ are given by
\ba
\delta _{(1)} P_{h}=-\frac{64\sigma}{3}\frac{\epsilon N}{k^3} \frac{H^2}{M_p^2} \left(1-3\cos ^2 \theta \right) \quad , \quad
\delta _{(2)} P_{h}=-\frac{32N^2\sigma ^2 \epsilon}{k^3 c_L^5}\frac{H^2}{M_p^2} \sin ^4 \theta .
\ea
As a result, $\delta P_h$ in total is 
\ba \label{Ph-total-FX}
\delta P_h =\frac{64}{256} N g_*^\zeta \epsilon^2 \bigg[ c_L^3 (1- 3 \cos^2 \theta) - \frac{3 N}{10}
g_*^\zeta \sin^4 \theta  \bigg] P_\zeta^{(0)} \, .
\ea

Finally, the scalar-tensor cross-correlation is 
\ba \label{P-zeta-h-FX}
P_{\zeta h} = -2 \sqrt2 \frac{N \sigma}{c_L^5 k^3} \frac{H^2}{M_P^2} \sin^2 \theta = -\frac{8 \sqrt2}{15} N \epsilon g_*^\zeta P_\zeta^{(0)} \sin^2 \theta  \, .
\ea
Again, depending on the sign of $g_{*}^\zeta$ this can be either a correlation or an anti-correlation. 
In addition the ratios $P_{\zeta h}/P_\zeta^{(0)}$ and $\delta P_h/P_\zeta^{(0)}$ do  not depend on
$F(X)$ explicitly, their dependence on $F(X)$ comes only indirectly via $\epsilon$.

The discussions of the contributions of the tensor anisotropies and their contributions in $TT$ correlation is the same as in previous model. We see  that  $\frac{P_{\zeta h}}{g_*^\zeta P_\zeta^{(0)}} \sim N \epsilon $ while   $\frac{\delta P_{ h}}{g_*^\zeta P_\zeta^{(0)}} \sim (N \epsilon)^2 g_*^\zeta$. As a result the contribution of $P_{\zeta h}$ in $TT$ correlation can be important while the contributions of  $\delta P_h$ is very small. Finally, the discussions of the IR
anisotropies are the same as in the previous model.
\\

\section{Statistical anisotropies on the CMB}

In this section we calculate the CMB anisotropies. The temperature fluctuation $\Delta T/T$ can be expanded under the bases of spherical harmonics with coefficients $a_{lm}$. The correlation of $a_{lm}$ can be calculated as
\begin{align}\label{eq:aXaX}
\langle a^{X_1}_{l_1, m_1} a^{X_2}_{l_2, m_2}\rangle = 4 \pi
\int \frac{dk}{k} \Delta_{l_1}^{i_1 X_1}(k) \Delta_{l_2}^{i_2 X_2}(k)
\int  d\Omega ~  [{}_{i_1}Y^*_{l_1m_1}(\theta, \phi)] [{}_{i_2}Y_{l_2m_2}(\theta, \phi)] P^{i_1, i_2}(k, \theta, \phi)~,
\end{align}
where $X^i$ takes value (T, E, B), which are the temperature anisotropy, the E-mode and B-mode respectively. Here we only consider anisotropy which has a rotational symmetry along the rotation of angle $\phi$. Thus the angular momentum along this direction is conserved and $\langle a^{X_1}_{l_1, m_1} a^{X_2}_{l_2, m_2}\rangle$ is non-vanishing only when $m_1 = m_2$. On the other hand, the rotation along the $\theta$ direction is no longer a symmetry of the system. Thus in addition to the diagonal correlations with $l_1=l_2$, there can also be non-vanishing correlations with $l_1=l_2\pm1$ and $l_1=l_2\pm3$ for TB and EB, and with $l_1=l_2\pm 2$ and $l_1=l_2\pm 4$ for TT, TE, EE and BB, respectively.

The $\Delta_{l}^{i X}(k)$ parts of \eqref{eq:aXaX} are the radiation transfer functions, which we compute using the public code of ``the Cosmic Linear Anisotropy Solving System'' (CLASS) \cite{Blas:2011rf}. The ${}_{i}Y_{lm}(\theta, \phi)$ in \eqref{eq:aXaX} denotes the spin-$i$-weighted spherical harmonics. With the spin-weighted basis, the $P^{i_1, i_2}$ can be calculated by
\begin{align}
  P^{0,0} = P_{\zeta}~, \quad P^{0,\pm2} = (P^{\pm2,0})^* = \frac{1}{\sqrt2} \left( P_{\zeta h_+} \pm i P_{\zeta h_\times}\right)~,
\end{align}
\begin{align}
  P^{\pm2, \pm2} = \frac{1}{2} \left( P_{h_+} + P_{h_\times}\right)~, \quad P^{\pm2, \mp2} = \frac{1}{2} \left( P_{h_+} - P_{h_\times}\right)~,
\end{align}
where $P_{\zeta h_\times}=0$ at leading order for our models, and the other power spectra $P_{\zeta}$, $P_{h_+}$, $P_{h_\times}$ and $P_{\zeta h_+}$ are calculated in Eqs. \eqref{delta-P-zeta2}, \eqref{Ph-total} and \eqref{P-zeta-h} for the original $F_Y, F_Z \sim F$ model and Eqs. \eqref{eq:delta-P-zeta-FX}, \eqref{Ph-total-FX} and \eqref{P-zeta-h-FX} for the $F(X)$ model respectively \footnote{Note that in \eqref{Ph-total} and \eqref{P-zeta-h}, the $\delta_{(1)}P_h$ has contribution half from $P_{h_+}$ and the other half from $P_{h_\times}$. While $\delta_{(2)}P_h$ has contribution from $P_{h_+}$ only}.

\begin{figure}[!htb]
  \centering
  \includegraphics[width=0.45\textwidth]{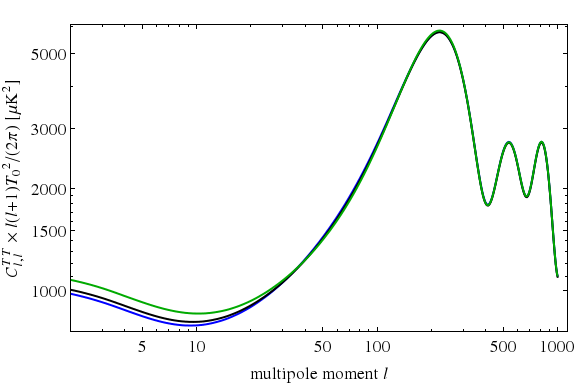}
  \hspace{0.05\textwidth}
  \includegraphics[width=0.45\textwidth]{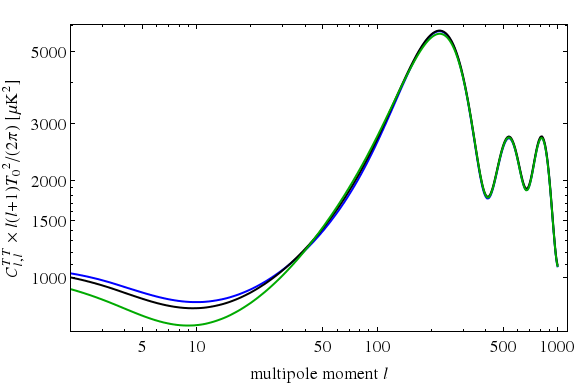}
  \caption{\label{fig:TT0} The TT correlation at $\ell_2=\ell_1$. The left panel is for $m=0$ and the right panel is for $m=\ell_1$. Here and hence after, the black curve represents the reference model with $g_*^\zeta = 0$. The blue line denotes the original solid inflation model with $F_Y, F_Z \sim F$, and the green line denotes the $F(X)$ model of solid inflation.}
\end{figure}

\begin{figure}[!htb]
  \centering
  \includegraphics[width=0.45\textwidth]{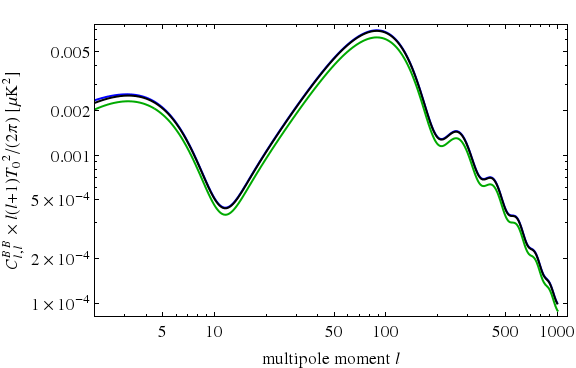}
  \hspace{0.05\textwidth}
  \includegraphics[width=0.45\textwidth]{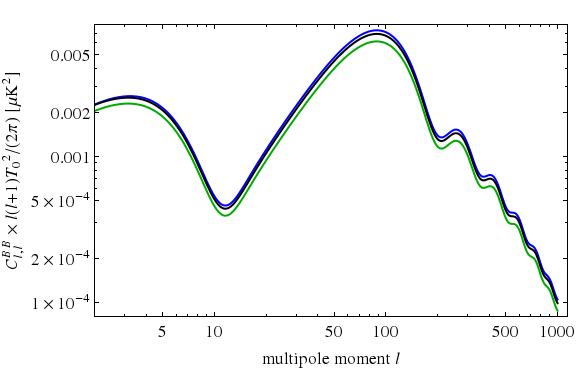}
  \caption{\label{fig:BB0} The $m=0$ (left) and $m=\ell_1$ (right) plots for BB correlation with $\ell_2=\ell_1$. }
\end{figure}

\begin{figure}[!htb]
  \centering
  \includegraphics[width=0.45\textwidth]{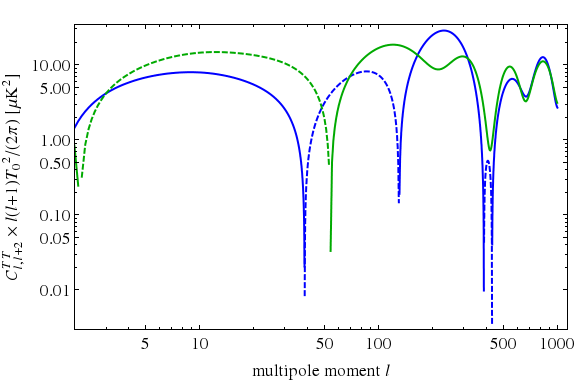}
  \hspace{0.05\textwidth}
  \includegraphics[width=0.45\textwidth]{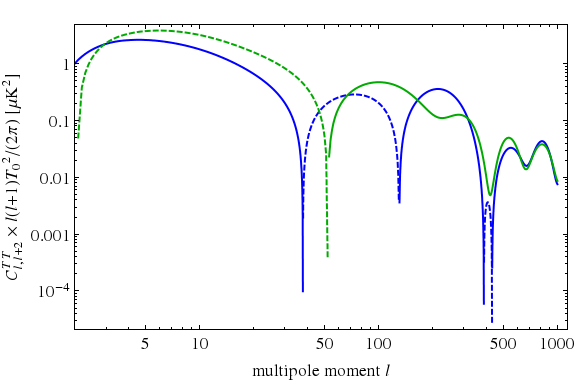}
  \caption{\label{fig:TT2} The $m=0$ (left) and $m=\ell_1$ (right) plots for TT correlation with $\ell_2=\ell_1+2$. Here and hence after, the dashed lines denote the plotted quantity (here $C_{l, l+2}^{TT}$) is negative along this line segment, and thus we plot $-C_{l, l+2}^{TT}$ on the logarithm scales.}
\end{figure}

\begin{figure}[!htb]
  \centering
  \includegraphics[width=0.45\textwidth]{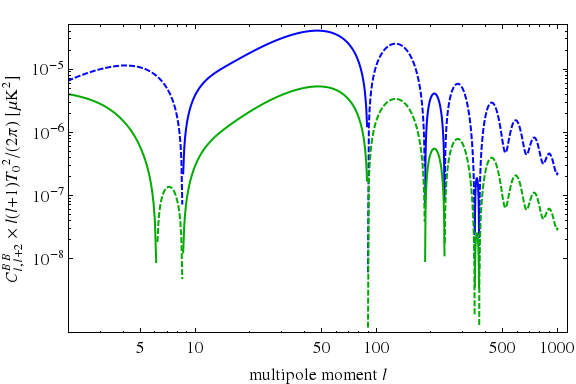}
  \hspace{0.05\textwidth}
  \includegraphics[width=0.45\textwidth]{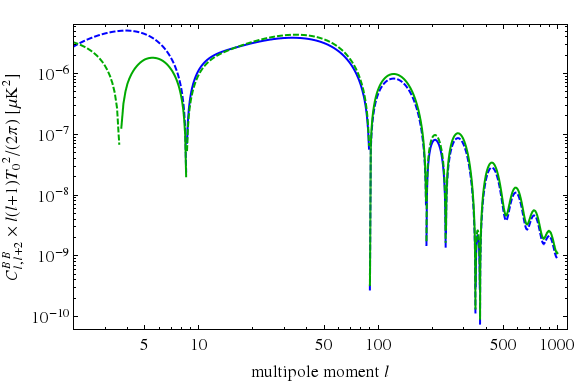}
  \caption{\label{fig:BB2} The $m=0$ (left) and $m=\ell_1$ (right) plots for BB correlation with $\ell_2=\ell_1+2$.}
\end{figure}

\begin{figure}[!htb]
  \centering
  \includegraphics[width=0.45\textwidth]{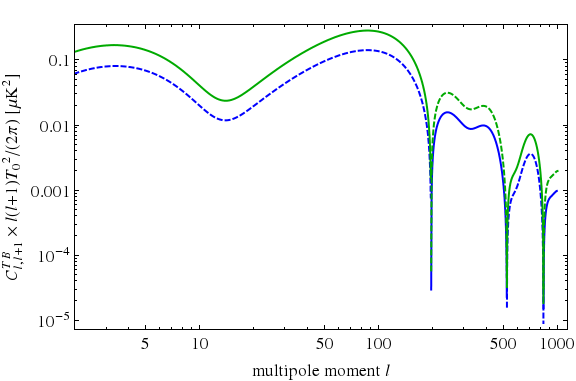}
  \hspace{0.05\textwidth}
  \includegraphics[width=0.45\textwidth]{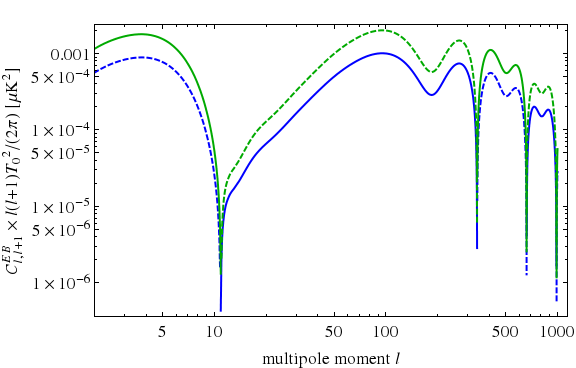}
  \caption{\label{fig:TBEB1} The $m=\ell_1$ plots for TB (left) and EB (right) correlation with $\ell_2=\ell_1+1$.}
\end{figure}

\begin{figure}[!htb]
  \centering
  \includegraphics[width=0.45\textwidth]{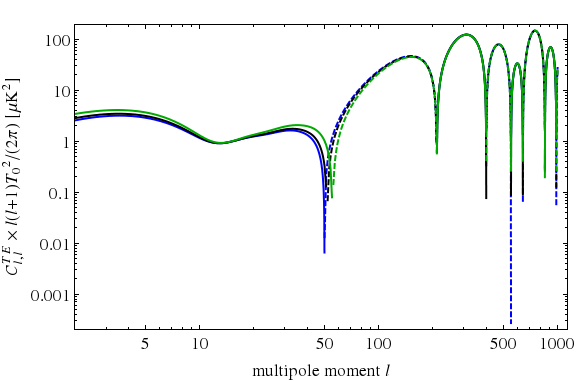}
  \hspace{0.05\textwidth}
  \includegraphics[width=0.45\textwidth]{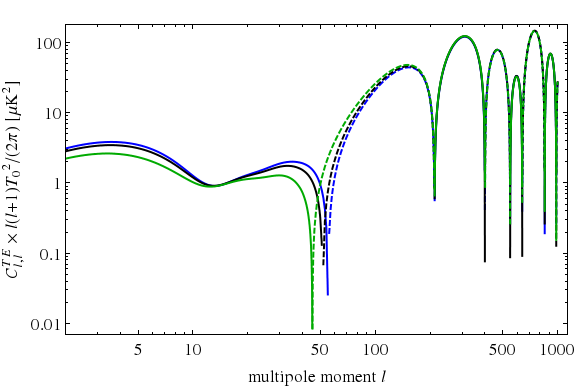}
  \caption{\label{fig:TE0} The $m=0$ (left) and $m=\ell_1$ (right) plots for TE correlation with $\ell_2=\ell_1$.}
\end{figure}

\begin{figure}[!htb]
  \centering
  \includegraphics[width=0.45\textwidth]{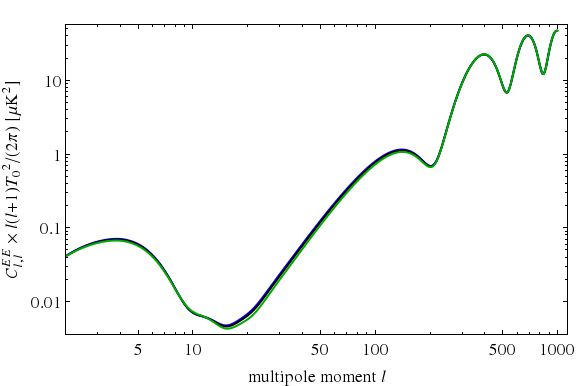}
  \hspace{0.05\textwidth}
  \includegraphics[width=0.45\textwidth]{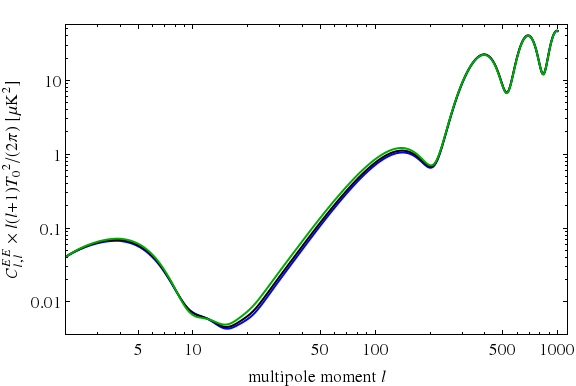}
  \caption{\label{fig:EE0} The $m=0$ (left) and $m=\ell_1$ (right) plots for EE correlation with $\ell_2=\ell_1$.}
\end{figure}

\begin{figure}[!htb]
  \centering
  \includegraphics[width=0.45\textwidth]{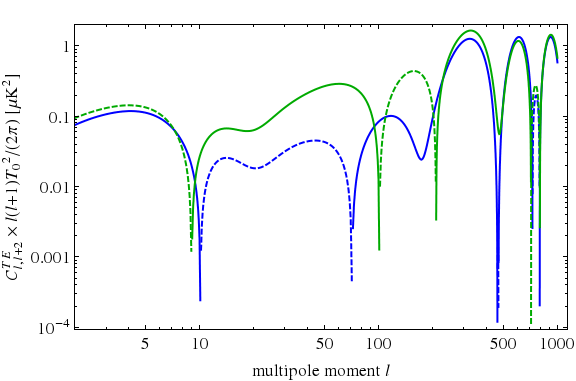}
  \hspace{0.05\textwidth}
  \includegraphics[width=0.45\textwidth]{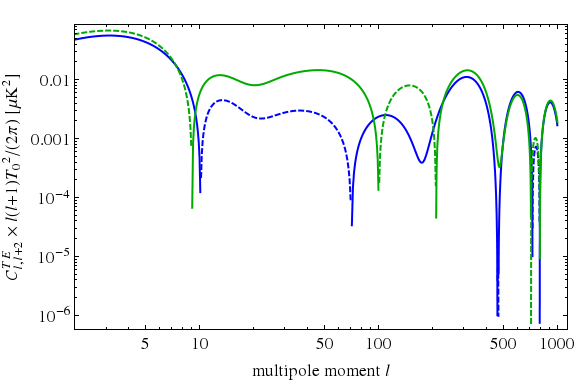}
  \caption{\label{fig:TE2} The $m=0$ (left) and $m=\ell_1$ (right) plots for TE correlation with $\ell_2=\ell_1+2$.}
\end{figure}

\begin{figure}[!htb]
  \centering
  \includegraphics[width=0.45\textwidth]{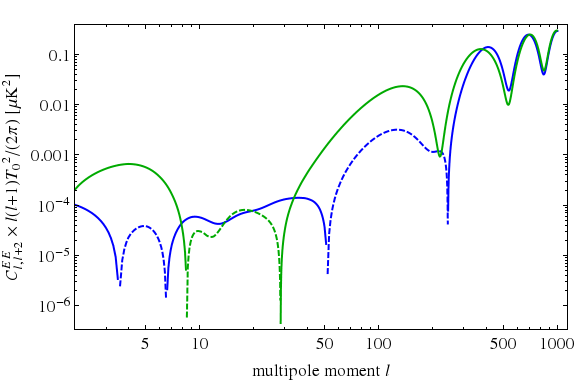}
  \hspace{0.05\textwidth}
  \includegraphics[width=0.45\textwidth]{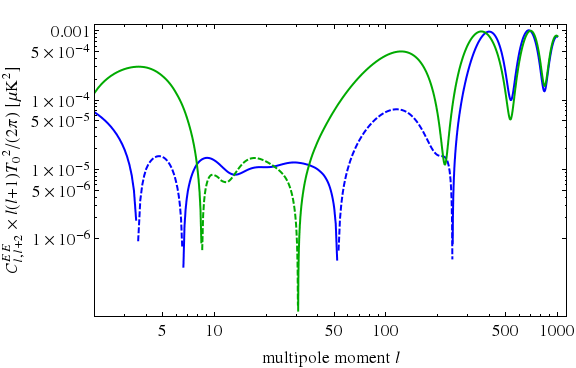}
  \caption{\label{fig:EE2} The $m=0$ (left) and $m=\ell_1$ (right) plots for EE correlation with $\ell_2=\ell_1+2$.}
\end{figure}

In Figs.~\ref{fig:TT0}-\ref{fig:EE2} the CMB observables are plotted. Note that now in $\langle a^{X_1}_{l_1, m} a^{X_2}_{l_2, m}\rangle$, $m$ should not be summed over for the purpose of probing anisotropies because otherwise some anisotropic signatures are averaged over in an uninteresting way. For illustration purpose, we plot the $m=0$ and $m=\min(\ell_1, \ell_2)$ values of those correlations. We plot the TT and BB power spectra in Figs.~\ref{fig:TT0}, \ref{fig:BB0}, \ref{fig:TT2} and \ref{fig:BB2}. In the anisotropic case, the TB and EB correlations are opened up, with $\ell_2 = \ell_1 \pm 1$. In Fig.~\ref{fig:TBEB1}, those cross-correlations are plotted. In Figs. \ref{fig:TE0}, \ref{fig:EE0}, \ref{fig:TE2} and \ref{fig:EE2}, the TE and EE correlations for $\ell_2=\ell_1$ and $\ell_2=\ell_1+2$ are plotted respectively.

\begin{figure}[!t]
  \includegraphics[width=0.45\textwidth]{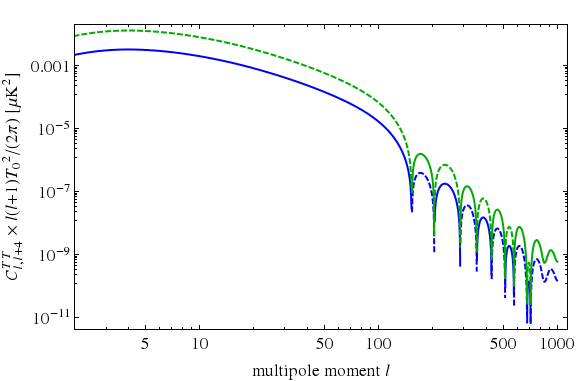}
  \hspace{0.05\textwidth}
  \includegraphics[width=0.45\textwidth]{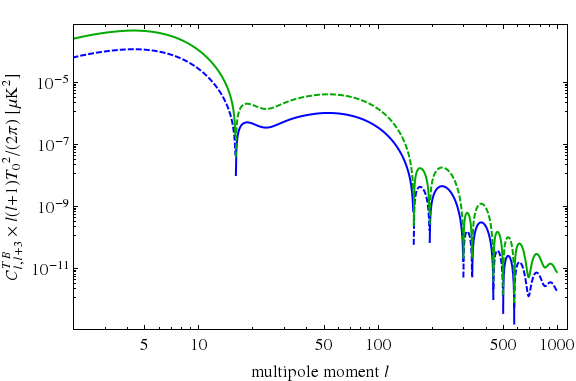}
  \caption{\label{fig:L3L4} The $m=L$ plots for TT correlation with $\ell_2=\ell_1+4$ (left) and TB correlation with $\ell_2=\ell_1+3$.}
\end{figure}

There are a few things that are interesting to observe from the plots:
\begin{itemize}
 
\item The scale dependence of the observed anisotropies are controlled by the tensor-to-scalar transfer function \cite{Chen:2014eua}. Considering that the tensor perturbations decay when they re-enter the horizon, the tensor-to-scalar transfer function decays towards large $k$. As a result, scale dependent anisotropies are generated even if the primordial anisotropies are isotropic. 

\item Relations between positive and negative $g_{*}^\zeta$ and the $\sin^4\theta$ term. We have only shown in the plots contribution from $g_{*}^\zeta$. For all the contributions except those from $\delta_{(2)}P_h$, the dependence on $g_{*}^\zeta$ is linear and thus the case of negative $g_{*}^\zeta$ is simply flipping the sign of corrections. However, this is not true for $\delta_{(2)}P_h$ (which results in the $\sin^4\theta$ term), which has quadratic dependence on $(g_{*}^\zeta)$. Because of the presence of the $\sin^4\theta$ term, now there are also TB and EB correlations for ($\ell, \ell+3$) and TT, TE, EE, BB correlations for ($\ell, \ell+4$), respectively. Here we do not  list all the plots, but instead show TT and TB for $m=\ell$ for illustration in Fig.~\ref{fig:L3L4}. 

\item The impact of cross-correlation. Unlike the case of \cite{Chen:2014eua}, here the impact of scalar-tensor cross-correlation plays a more significant role in the anisotropies. As a result, one can observe from Figs.~\ref{fig:TT0} that for some values of $m$, the low $\ell$ power spectrum is suppressed by the cross-correlation.  In \cite{Chen:2014eua, Zibin:2014iea, Emami:2014xga, Contaldi:2014zua} it has been pointed out that to calculate the power spectrum theoretically, one sums over $m$. After summing over $m$, the impact from the scalar-tensor cross-correlation cancels out and has no import on the temperature power spectrum. Nevertheless, the contribution exists for each $m$ and is interesting to study under less coarse graining of data.

\end{itemize}

\section{Conclusion}


In this paper we have studied statistical anisotropies in the model of anisotropic solid inflation with the particular emphasis on anisotropies in GW. As we discussed, solid inflation is not efficient in erasing anisotropic deformation of the background geometry so any anisotropy induced from the initial condition will persist for a long period during inflation. As a result, solid inflation provides a natural setup for anisotropic inflation. The recent detection of B-mode polarization by BICEP2 observation has opened the possibility of looking at B-mode polarization with non-trivial features such as the statistical anisotropies. The motivation of this work was to investigate whether observable statistical anisotropies can be imprinted on the CMB map from the quadrupole and octopole anisotropies in
scalar and the tensor power spectra and their cross-correlation.

We have calculated the anisotropies in scalar power spectrum $\delta P_\zeta$, the tensor power spectrum $\delta P_h$ and their cross-correlation $P_{\zeta h}$. Our expression for $\delta P_\zeta$ agrees exactly with the result obtained previously from a different method. We have shown that depending on the value of $g_*^\zeta$ the scalar-tensor 
can be either correlated or anti-correlated. In addition, the contribution of $P_{\zeta h}$ in $TT$
correlation compared to $\delta P_{\zeta}$ is at the order of $N \epsilon $. As a result, the scalar tensor correlation can have significant contribution in the effective value of $g_*$ in the $TT$ power anisotropies. However, the contribution of the tensor power spectra $\delta P_h$ is too small to affect the effective 
value of $g_*$.

It is informative to compare the impact on the CMB between our current calculation and the model of anisotropic inflation \cite{Chen:2014eua}. For example in
\cite{Chen:2014eua} it is found that the TB correlation is a sensitive probe of the gauge coupling.
Similarly, the TB correlation is sensitive to the anisotropy parameter $\sigma$. However, it is useful to note that in the case of solid inflation, cross-correlation of scalar and tensor dominates the tensor-related statistical anisotropies, which gives different prediction to the $TT$ power spectrum from that of the anisotropic inflation. In the former, the anisotropic correction to the $TT$ power spectrum changes its sign from $m=0$ to $|m|=\ell$, while in the latter the anisotropic correction keeps positive.

It is known that large non-Gaussianities can be generated in solid model. We have found that
the anisotropies in tensor perturbations and the cross-correlation have the same angular dependence as the bispectrum while 
the anisotropies in the tensor power spectrum have different angular dependence.
 
While $\delta  P_\zeta$ is insensitive to the duration of inflation, both $\delta P_h$ and $P_{\zeta h}$
depend on $N$, the total number of e-folds. If anisotropic solid inflation is prolonged in the past, the IR anisotropies accumulate \cite{Bartolo:2012sd}. The IR anisotropies are balanced by the slowly decaying behavior of anisotropies, which is described by Eq. \eqref{sigma-t}. Thus the IR mode of the anisotropies are allowed to grow for of order $3/(4c_T^2 \epsilon)$ e-folds before the exponential decay behavior shuts off the IR growth. Such a balance should set a characteristic amount of anisotropy for solid inflation, assuming a long period of solid inflation. 

Finally, we would like to mention a caveat about the vanishing correction from the tensor-scalar cross-correlation to the $m$ averaged power spectrum. Note that the actually observed map has been masked against foreground, the unmasked regime (for example for the CL31 mask only 31.71\% of the sky is retained \cite{Ade:2013kta}) may not represent an efficient average against $m$. It would be interesting to check if the effect of the mask may reopen the possibility for low $\ell$ suppression from anisotropies, which has the potential to reconcile the tension between Planck and BICEP2.

\section*{Acknowledgment} 
We thank A. A. Abolhasani, N. Bartolo, X. Chen, S. Matarrese, M. H. Namjoo, 
A. Ricciardone, M. Sasaki  
and  M. Zarei for helpful discussions. YW is supported by a Starting Grant of the European Research Council (ERC STG grant 279617), and the Stephen Hawking Advanced Fellowship.



\end{document}